\documentclass[a4paper,11pt]{article}
\usepackage{jheppub} % for details on the use of the package, please see the JINST-author-manual
\usepackage{overpic}
\usepackage{cleveref}
\usepackage{subfig}
\usepackage{subcaption} % replace \usepackage{subfigure}
\usepackage{float} 
\usepackage{siunitx}
\usepackage{graphicx}
\usepackage{amsmath}
\title{\boldmath Measurement of the hyperon weak radiative decay $\Xi^0\to\gamma\Sigma^0$ at BESIII}

\collaboration{The BESIII Collaboration}
\collaborationImg{\includegraphics[height=30mm,angle=90]{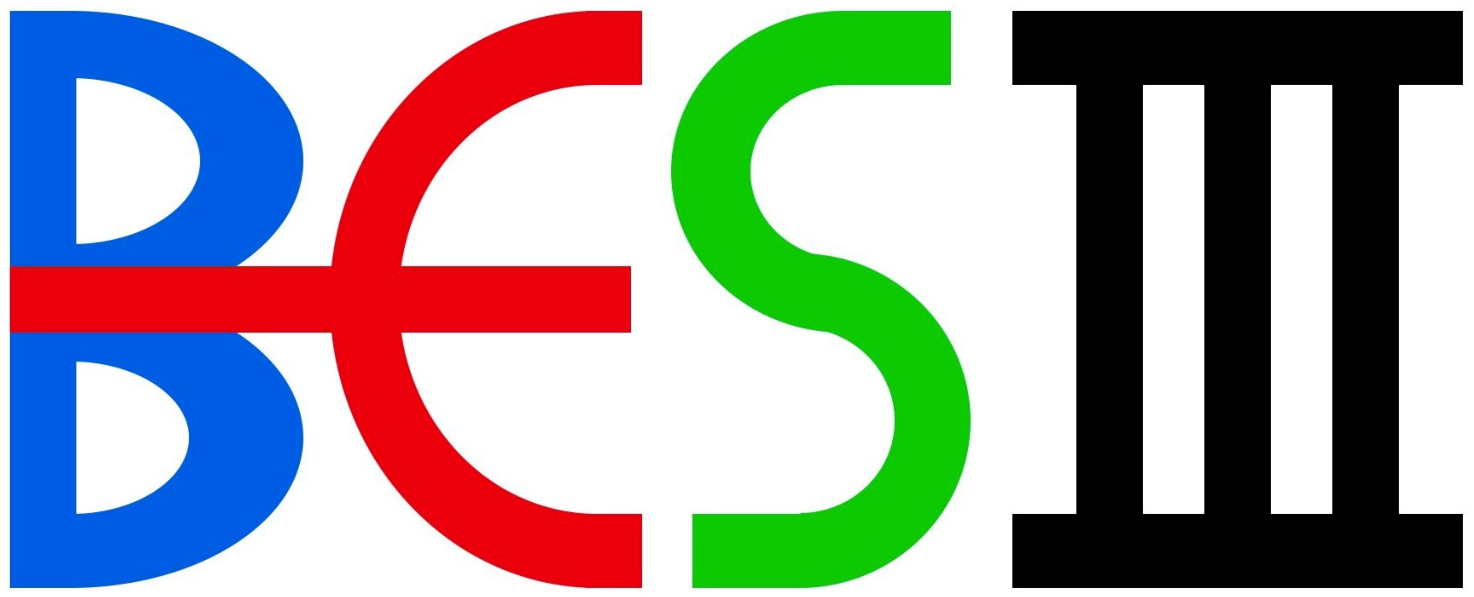}}

\emailAdd{besiii-publications@ihep.ac.cn}

\begin{document}

\abstract{The hyperon weak radiative decay $\Xi^0\to\gamma\Sigma^0$ is measured via the process $J/\psi\to\Xi^0\bar\Xi^0$. The absolute branching fraction of $\Xi^0\to\gamma\Sigma^0$  is determined to be $(3.69\pm 0.21_{\text{stat}}\pm0.12_{\text{syst}})\times 10^{-3}$, based on $(10.087\pm 0.044)\times 10^{9}$ $J/\psi$ events collected with the BESIII detector operating at the BEPCII collider. The decay asymmetry parameter is measured, with a complete angular analysis of the cascade decay chain, to be $\alpha_{\gamma} = -0.807\pm 0.095_{\text{stat}}\pm 0.011_{\text{syst}}$.}

\keywords{$e^{+}e^{-}$ Experiments, Hyperon, Weak Radiative Decay}

\newcommand{\BESIIIorcid}[1]{\href{https://orcid.org/#1}{\hspace*{0.1em}\raisebox{-0.45ex}{\includegraphics[width=1em]{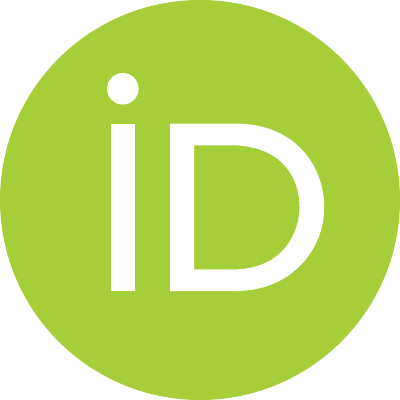}}}}

\maketitle
\flushbottom

\section{\boldmath Introduction}
\label{sec:intro}
The decay properties of hyperons have attracted great attention since their discovery in the 1950s~\cite{PhysRev.80.1099}~\cite{PhysRevLett.2.215}.  Among them, the hyperon weak radiative decay (HWRD) is considered the simplest because it is a two-body decay and free from strong interactions among the final states. Both parity-conserving and parity-violating amplitudes contribute to the weak decay of hyperons, which can be quantified with the decay parameters. In 1964, Yasuo Hara predicted that the $\Sigma^+\to\gamma p$ and $\Xi^-\to\gamma\Sigma^-$ decay asymmetry parameters should be zero as the parity-violating amplitude should vanish in the $\mathrm{SU}(3)$ symmetry limit (Hara's theorem)~\cite{PhysRevLett.12.378}. The decay asymmetry parameter is given by 
\begin{equation}
    \alpha_{\gamma}= \frac{2\mathrm{Re}(a b ^{*})}{|a|^2+|b|^2},
\end{equation}
where $a$ and $b$ are the parity-conserving and parity-violating amplitudes, respectively.
In contrast to the prediction of this theory, a large and negative decay asymmetry parameter in the $\Sigma^+\to \gamma p$ decay was first observed at Berkeley~\cite{PhysRev.188.2077} and later confirmed by KEK~\cite{PhysRevLett.59.868}, Fermilab~\cite{PhysRevLett.68.3004} and recently BESIII~\cite{PhysRevLett.130.211901}. The unsuccessful explanation of these experimental results using the SU(3) approach was regarded as one of the long-standing puzzles in the low four-momentum transfer region.

There are six HWRD processes involving the transition $s\to d\gamma$ for spin-$1/2$ baryons. Their branching fractions and decay asymmetry parameters have all been measured except for the process $\Sigma^0\to\gamma n$, which is expected to be overwhelmed by the radiative decay $\Sigma^0\to\gamma\Lambda$. The experimental and theoretical results of the other five HWRD processes are summarized in table~\ref{tab:allandBES}. Most results ~\cite{PhysRevLett.63.2717, NA48, PhysRevLett.86.3239, 2010241} were obtained at fixed-target experiments using hyperon beams, before BESIII.  Using pair production of hyperons and anti-hyperons with non-zero transverse polarization, BESIII has great potential to study HWRD processes. Many phenomenological models~\cite{PhysRevD.73.076005, CPCabc067, PhysRevD.59.054019, NARDULLI1987187, GAVELA1981417} have been proposed to explain these HWRD decays. However, there is no theoretical model that can fully describe them. In addition, BESIII has measured the HWRD processes $\Lambda \to \gamma n $~\cite{PhysRevLett.129.212002}, $\Sigma^+ \to \gamma p$~\cite{PhysRevLett.130.211901} and $\Xi^0\to\gamma\Lambda$~\cite{BAM760}. In the $\Lambda \to \gamma n $~\cite{PhysRevLett.129.212002}  and $\Sigma^+ \to \gamma p$~\cite{PhysRevLett.130.211901} analyses, both the measured decay asymmetry parameters are the most precise to date and the measured branching fraction, $\mathcal{B}$, deviates significantly from the previous measurements by $5.6 \sigma$ and $4.2 \sigma$, respectively. To discriminate between the different theoretical approaches and to understand the difference in the HWRD process measurements between fixed target and $e^+e^-$ collision experiments, systematic measurements of all HWRDs with high precision are needed.

%There are six typical HWRD processes for $\mathrm{SU}(3)$ baryons. Their results on the branching fractions and decay asymmetry parameters have been measured experimentally except the process $\Sigma^0\to\gamma n$, that is supposed to be overwhelmed by the radiative decay $\Sigma^0\to\gamma\Lambda$. Most experiments were performed at a fixed target with hyperon beam facility before BESIII. With the pair production of hyperon anti-hyperon with non-zero transverse polarization, BESIII shows great potential to study of the HWRD processes. BESIII has measured HWRD $\Lambda \to n \gamma$~\cite{PhysRevLett.129.212002}, $\Sigma^+ \to p\gamma$~\cite{PhysRevLett.130.211901} and $\Xi^0\to\Lambda \gamma$~\cite{BAM760}. In $\Lambda \to n \gamma$~\cite{PhysRevLett.129.212002}  and $\Sigma^+ \to p\gamma$~\cite{PhysRevLett.130.211901} analysis, both measured decay asymmetry parameters are the most precise to date and measured branching fraction deviate from previous measurement significantly, to be $5.6 \sigma,\ 4.2 \sigma$ respectively, which may indicate more experimental results obtained from fixed-target experiments should be checked on electron-positron collision experiments.

%To address the conflicts between theory and experiment, many phenomenological models have been proposed. The experimental and theoretical results are summarized in Table~\ref{tab:alltheory_BR} and Table~\ref{tab:alltheory}. For discrimination among different theoretical approaches, measurements of all HWRD with high precision are needed.

\begin{table}[htbp]
    \centering
    \resizebox{\linewidth}{!}{
    \begin{tabular}{c|ccccc}
    \hline
     Decay    & $\Lambda \to \gamma n$ & $\Sigma^{+} \to \gamma p$ &$\Xi^0 \to \gamma \Lambda$ & $\Xi^0 \to \gamma \Sigma^0$ & $\Xi^-\to\gamma\Sigma^-$ \\
     \hline \multicolumn{6}{c}{ Branching fraction, $\mathcal{B}$, in units of $\times 10^{-3}$ } \\
     \hline
      VMD~\cite{PhysRevD.73.076005}   & 0.77 & 0.72 & 1.02 & 4.42 &0.16\\
      NRCQM~\cite{CPCabc067} &$1.83 \pm 0.96$ & $1.06 \pm 0.59$ & $0.96 \pm 0.32$ & $9.75 \pm 4.18$ &$-$\\
      ChPT~\cite{PhysRevD.59.054019} & 1.45 & 16 & 1.17 & 1.14 & $-$\\
      Pole model(I)~\cite{NARDULLI1987187} & $0.16 \pm 0.06$ & $0.75 \pm 0.3$ & $0.72 \pm 0.42$ & $2.6 \pm 1.2$  & 0.04\\
      Pole model(II)~\cite{GAVELA1981417} & 0.62 & 1.15 & 3.0 & 7.2 & $-$\\
      PDG~\cite{PDG2018} & $1.75 \pm 0.15$  & $1.23 \pm 0.05$ & $1.17 \pm 0.07$ & $3.33 \pm 0.10$ & 0.127\\
      BESIII~\cite{PhysRevLett.129.212002}~\cite{PhysRevLett.130.211901}~\cite{BAM760} & $0.832\pm0.07$ & $0.996\pm0.028$ & $1.347\pm0.085$ & this work & $-$\\
      \hline \multicolumn{6}{c}{ Decay asymmetry parameter, $\alpha_{\gamma}$ } \\
      \hline
      VMD~\cite{PhysRevD.73.076005}   & $-0.93$ & $-0.67$ & $-0.97$ & $-0.92$ &0.89\\
      NRCQM~\cite{CPCabc067} &$-0.67 \pm 0.06$ & $-0.58 \pm 0.06$ & $0.72 \pm 0.11$ & $0.33 \pm 0.036$ &$-$\\
      ChPT~\cite{PhysRevD.59.054019} & $-0.19$ & $-0.49$ & 0.46 & 0.15 & 0.84\\
      Pole model(I)~\cite{NARDULLI1987187} & 0.91 & $-0.92$ & 0.07 & $-0.75$  & 0\\
      Pole model(II)~\cite{GAVELA1981417} & $-0.49$ & $-0.80$ & $-0.78$ & $-0.96$ & $-$\\
      PDG~\cite{PDG2018} &  $-$  & $-0.76 \pm 0.08$ &  $-0.70 \pm 0.07$ & $-0.69 \pm 0.06$ & $-$\\
      BESIII~\cite{PhysRevLett.129.212002}~\cite{PhysRevLett.130.211901}~\cite{BAM760} & $-0.16\pm0.11$ & $-0.652\pm0.059$ & $-0.741\pm0.065$ & this work & $-$\\
      \hline
    \end{tabular}
    }
    \caption{Theoretical predictions and experimental measurements of HWRD branching fractions and decay asymmetry parameters.  The cited PDG averages do not include later BESIII results.}
    \label{tab:allandBES}
\end{table}

For the process $\Xi^0\to \gamma\Sigma^0$, the existing experimental results are summarized in table~\ref{tab:exp}. None of the phenomenological models are fully satisfying. The measured branching fractions for $\Xi^0\to\gamma\Sigma^0$ use $\mathcal B (\Xi^0\to\Lambda\pi^0)$ as a normalization, and possible $CP$ asymmetries are rarely discussed. In addition, the radiative decay asymmetry parameter requires input from the $\Lambda\to p\pi^{-}$ decay, where a recent BESIII result has corrected the long-standing decay asymmetry by 17\%\cite{PhysRevLett.129.131801}. The $\Xi^{0}$ is the only hyperon that possesses two different HWRD decays. Comparing these two processes will shed light on the mystery of the spin composition of hyperons. The octet members $\Sigma^0$ and $\Lambda$ represent orthogonal three-quark states with different U-spins. Therefore, the studies of the radiative decays $\Xi^0 \to\gamma \Lambda$ and $\Xi^0 \to \gamma\Sigma^0 $ give information about the hyperon structure~\cite{NA48,Shi:2025xkp}. The updated results on $\Xi^{0}\to\gamma\Lambda$~\cite{BAM760} and the results from the analysis presented in his paper provide an updated ratio of the two branching fractions and can be used to discriminate between various theoretical models~\cite{PhysRevD.73.076005, CPCabc067, PhysRevD.59.054019, NARDULLI1987187, GAVELA1981417}.
%~\cite{PhysRevD.73.076005}~\cite{CPCabc067}~\cite{PhysRevD.59.054019}~\cite{NARDULLI1987187}~\cite{GAVELA1981417}.

\begin{table}[htbp]
    \centering
    \resizebox{\linewidth}{!}{
    \begin{tabular}{c|c|c|c}
        \hline
        Experiment & $\mathcal{B}$ ($\times 10^{-3}$) & Decay asymmetry, $\alpha_\gamma$ & Events \\
        \hline
        1989 SPEC~\cite{PhysRevLett.63.2717} & $3.56\pm 0.42\pm 0.10$ & $+0.20\pm 0.32\pm 0.05$  & $85$\\
        2000 NA48~\cite{NA48} & $3.16\pm 0.76 \pm 0.32$ & $-$ & $17$ \\   
        2001 KTeV~\cite{PhysRevLett.86.3239} & $3.34\pm 0.05 \pm 0.09$ & $-0.63\pm 0.08 \pm 0.05$ & $4045$ \\
        2010 NA48~\cite{2010241} & $-$ & $-0.729\pm0.030\pm0.076$ & 15 K \\
        2010 NA48~\cite{2010241}(c.c. channel) & $-$ & $+0.786\pm0.104_{\text{stat}}$ & 1404 \\
        \hline
    \end{tabular}
    }
    \caption{Experimental results of the branching fractions and decay asymmetry parameters of $\Xi^0 \to \gamma\Sigma^0$. The first uncertainties are statistical and the second are systematic.}
    \label{tab:exp}
\end{table}

In this paper, we report on improved absolute branching fraction and decay asymmetry measurements of $\Xi^0\to\gamma\Sigma^0$ and its charge conjugation (c.c.) based on $(10.087\pm 0.044)\times 10^{9}\ J/\psi$ events collected with the BESIII detector~\cite{BESIII} at the BEPCII collider~\cite{BEPCII}.

\section{BESIII detector and Monte Carlo simulation}
\label{sec:detector}

The BESIII detector records symmetric $e^+e^-$ collisions provided by the BEPCII storage ring, which operates with a peak luminosity of $1.1\times10^{33}$~cm$^{-2}$s$^{-1}$ in the center-of-mass energy range from 1.84 to 4.95~GeV. BESIII has collected large data samples in this energy region~\cite{Ablikim_2020}. The cylindrical core of the BESIII detector covers 93\% of the full solid angle and consists of a helium-based multilayer drift chamber~(MDC), a plastic scintillator time-of-flight system~(TOF), and a CsI(Tl) electromagnetic calorimeter~(EMC), which are all enclosed in a superconducting solenoidal magnet providing a 1.0~T magnetic field. The solenoid is supported by an octagonal flux-return yoke with resistive plate counter muon identification modules interleaved with steel. The charged-particle momentum resolution at $1~{\rm GeV}/c$ is $0.5\%$, and the d$E$/d$x$ resolution is $6\%$ for electrons from Bhabha scattering. The EMC measures photon energies with a resolution of $2.5\%$ ($5\%$) at $1$~GeV in the barrel (end cap) region. The time resolution in the TOF barrel region is 68~ps. The end cap TOF system was upgraded in 2015 using multi-gap resistive plate chamber technology, providing a time resolution of 60~ps~\cite{etof}. About 87\% of the data used in this paper benefits from this upgrade.  

\begin{figure}[htbp]
    \centering
    \includegraphics[width=0.8\textwidth]{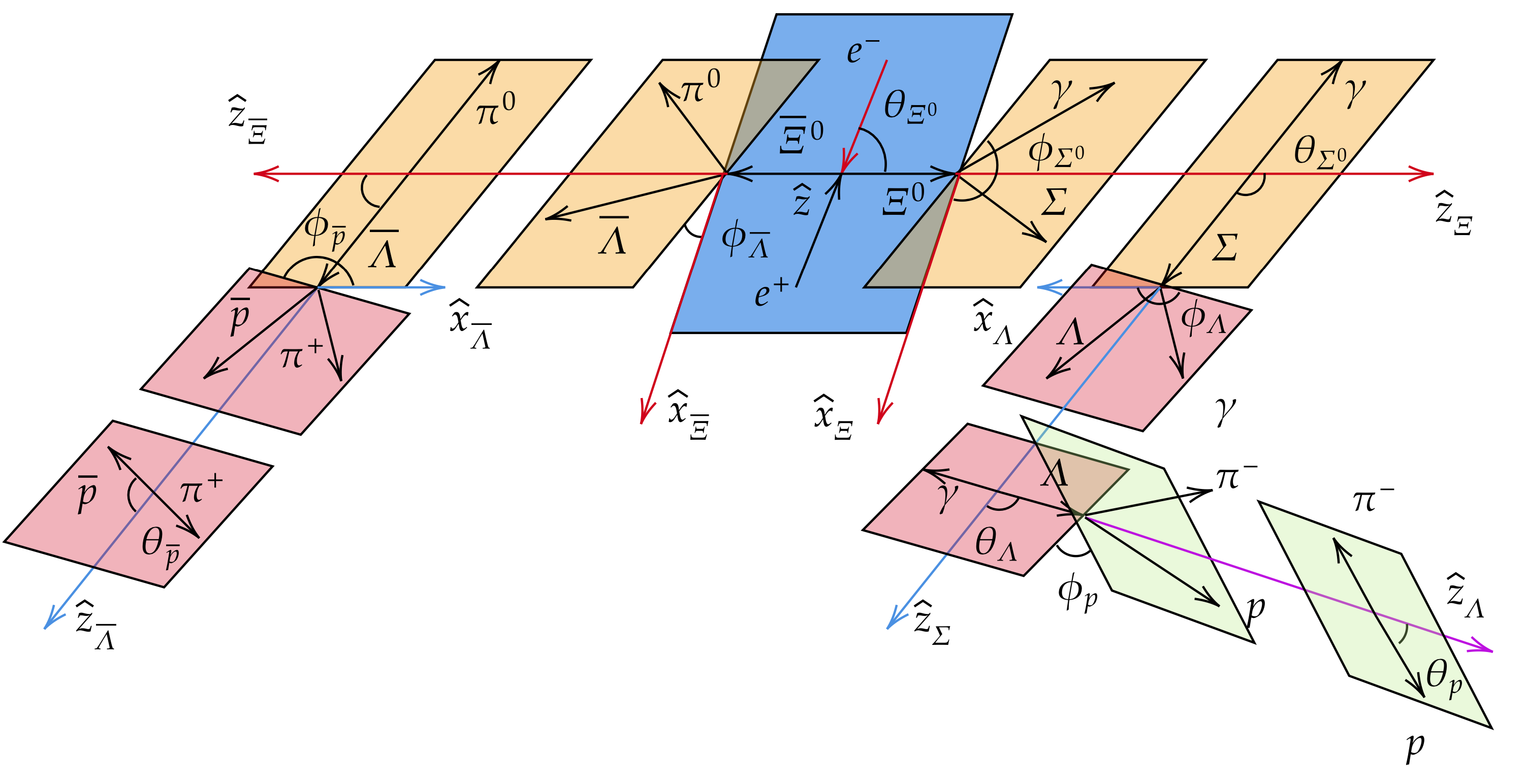}
    \caption{The helicity frame and the helicity angle in the decay of $J/\psi\to\Xi^0(\to\gamma\Sigma^0)\bar\Xi^0(\to\bar\Lambda\pi^0)$.}
    \label{fig:decaytopology}
\end{figure}

Monte Carlo (MC) simulated events are used to determine the detection efficiency, optimize selection criteria, and study possible backgrounds. GEANT4-based MC simulation software, which includes the geometric and material descriptions of the BESIII detector, the detector response, and digitization models, as well as the detector running conditions and performance, is used to generate MC samples. The simulation models the beam energy spread and initial state radiation in the $e^+e^-$ annihilations with the generator {\sc KKMC}~\cite{kkmc}. The inclusive MC sample includes the production of the $J/\psi$ resonance and the continuum processes incorporated in {\sc KKMC}. The known decay modes are modeled with {\sc EvtGen}~\cite{LANGE2001152} using branching fractions taken from the Particle Data Group~\cite{PDG}, and the remaining unknown charmonium decays are modeled with {\sc LundCharm}~\cite{Yang_2014}. Final state radiation from charged final state particles is incorporated using the {\sc PHOTOS} package~\cite{PHOTOS1, PHOTOS2, PHOTOS3}. 
A general formula for the differential cross sections of the signal $J/\psi\to\Xi^0(\to\gamma\Sigma^0) \, \bar\Xi^0(\to\bar\Lambda\pi^0)$ contains eleven measurable angles $\xi=(\theta_{\Xi^0}, \theta_{\Sigma^0},\phi_{\Sigma^0},\theta_{\Lambda}$, $\phi_{\Lambda}, \theta_p, \phi_p, \theta_{\bar{\Lambda}}, \phi_{\bar{\Lambda}}, \theta_{\bar{p}}, \phi_{\bar{p}})$, where $\theta_{\Xi^0}$ is the polar angle of $\Xi^0$ in the $J/\psi$ rest frame; $\theta_{\Sigma^0},\phi_{\Sigma^0}$ are the polar and azimuthal angles of $\Sigma^0$ with respect to $\Xi^0$ helicity frame, respectively; $\theta_{\Lambda}\left(\theta_{\bar{\Lambda}}\right)$ and $\phi_{\Lambda}\left(\phi_{\bar{\Lambda}}\right)$ are the polar and azimuthal angles of $\Lambda(\bar{\Lambda})$ with respect to $\Sigma^0\left(\bar{\Xi}^0\right)$ helicity frame, respectively; and $\theta_p\left(\theta_{\bar{p}}\right)$ and $\phi_p\left(\phi_{\bar{p}}\right)$ are the polar and azimuthal angles of $p(\bar{p})$ with respect to $\Lambda(\bar{\Lambda})$ helicity frame, respectively. The definition of the helicity frames as well as the helicity angles above are illustrated in figure~\ref{fig:decaytopology}. With these helicity angles, the differential cross section can be constructed under the helicity formalism: $\mathrm d \sigma \varpropto \mathcal{W}(\xi) \, \mathrm d \xi$, where
\begin{equation}
    \mathcal{W}=
    \sum_{\mu, \nu, \mu', \nu', \rho=0}^3                     
    \, C_{\mu \nu}             
    \, a_{\mu \mu'}^{\Xi}      
    \, a_{\nu \nu'}^{\bar{\Xi}} 
    \, a_{\mu'\rho}^{\Sigma} 
    \, a_{\nu'0}^{\bar\Lambda}  
    \, a_{\rho0}^{\Lambda}.
    \label{eq:decay_amplitude}
\end{equation}
Here $C$ represents the polarization and spin correlation matrix of $\Xi^0\bar\Xi^0$, and $a$ are the decay matrices of the hyperons, which depend on the decay products~\cite{amplitude}. The input decay parameters for the simulation are fixed to those of the latest measurement~\cite{PhysRevD.108.L031106} except for $\alpha_{\Xi^0\to\gamma\Sigma^0}$, which is determined in this analysis. A signal MC sample of $1.58$ million events is generated, corresponding to 100 times the signal in the data.

\section{Event selection and background analysis}
\label{sec:bkgana}
A double-tag technique is used to select the $e^+e^-\to J/\psi\to\Xi^0 \, \bar\Xi^0$ candidate events, where a $\bar\Xi^0$ is first tagged by $\bar\Xi^0\to \bar\Lambda \pi^0$, a single-tag (ST), and the signal $\Xi^0 \to\gamma\Sigma^0$ is selected from among the particles recoiling against the ST $\bar\Xi^0$. The combination of a ST and a signal candidate is a double-tag (DT).  The c.c.\ mode is always implied throughout this paper, unless explicitly stated.

The charged tracks detected in the MDC are required to be within a polar angle range of $|\cos\theta| < 0.93$, where $\theta$ is defined with respect to the $z$-axis, which is the symmetry axis of the MDC. In order to identify charged particles, a likelihood-based particle identification (PID) method is employed. This method combines measurements of the specific ionization energy loss in the MDC ($\mathrm d E/\mathrm d x$) and the time of flight in the TOF to form likelihoods $\mathcal L(h) \, (h = p, K, \pi)$ for each hadron, $h$, hypothesis. Tracks are identified as protons when $\mathcal L(p) > \mathcal L(K)$, $\mathcal L(p) > \mathcal L(\pi)$ and the momentum is larger than $0.3\ \mathrm{GeV}/{\it c}$. Tracks other than protons are assigned as pions without PID criteria to improve the detection efficiency. At least one $\bar p$ and one $\pi^+$ are required for ST selection. At least one $p$ and one $\pi^-$ are further required for DT selection.

To reconstruct the $\bar\Lambda$, a vertex fit is performed on the $\bar p\pi^+$ combinations. The $\bar\Lambda$ candidates are required to satisfy $|M_{\bar p\pi^+} - M_{\bar\Lambda}| < 6 \ \mathrm{MeV}/{\it c}^2$, where $M_{\bar\Lambda}$ is the $\bar\Lambda$ nominal mass~\cite{PDG} and $M_{\bar p\pi^+}$ is the fitted invariant mass of the $\bar p \pi^+$. 

Photon candidates are reconstructed from isolated showers in the EMC.  To veto secondary showers produced by interaction between charged tracks and detector material, the isolation criteria are separation from anti-protons by at least 20 degrees, or 10 degrees for other charged tracks.  
Each photon candidate is required to have a minimum energy of $25\ \mathrm{MeV}$ in the EMC barrel region with polar angle satisfying $|\cos\theta|<0.80$ or $50\ \mathrm{MeV}$ in the end-cap region with $0.86<|\cos\theta|<0.92$. To improve the reconstruction efficiency and the energy resolution, the energy deposited in the nearby TOF counters is included in the photon reconstruction. To suppress electronic noise and showers unrelated to the event, the difference between the EMC time and the event start time is required to be within $[0, 700]\ \mathrm{ns}$. The $\pi^0$ candidates are reconstructed by requiring the invariant mass of photon pairs ($M_{\gamma \gamma}$) to satisfy $115 < M_{\gamma\gamma} < 150\ \mathrm{MeV}/{\it c}^{2}$. The asymmetric mass window accounts for energy leakage out of the EMC. The momenta of the $\pi^0$ are updated with a one-constraint fit to the nominal mass~\cite{PDG}. At least one $\pi^0$ is required on the ST side, and two additional photons are required for the DT selection. 

The ST $\bar\Xi^0$ candidate is reconstructed with the $\bar\Lambda\pi^0$ combination whose invariant mass $M_{\bar\Lambda\pi^0}-M_{\bar p \pi^+}+M_{\bar\Lambda}$ is the closest to the nominal $\Xi^0$ mass, $M_{\Xi^0}$,~\cite{PDG}, and is required to be within $|M_{\bar\Lambda\pi^0}-M_{\bar p \pi^+}+M_{\bar\Lambda}-M_{\Xi^0}|<12\ \mathrm{MeV}/{\it c}^2$. The ST yield is extracted by performing a binned maximum-likelihood fit to the recoil mass distribution, defined as $M_{\text{rec}}=\sqrt{(E_{\text{cms}}-E_{\bar p\pi^+}-E_{\pi^0})^2-(\mathbf p_{\bar p\pi^+}+\mathbf p_{\pi^0})^2}$, where $E_{\text{cms}}$ is the center-of-mass energy, and $E_{\bar p\pi^+}(\mathbf p _{\bar p \pi^+})$ and $E_{\pi^0} (\mathbf p_{\pi^0})$ are the energies (momenta) of the $\bar\Lambda$ and the $\pi^0$ in the $J/\psi$ rest frame, respectively.  The fit shape for $M_{\text {rec}}$ is obtained from MC signal events with the additional requirement that the reconstructed and MC-generated angles of the $\pi^0$ agree within $20^{\circ}$. 

\begin{figure}[htbp]
    \centering
    \begin{overpic}[width=0.48\textwidth]{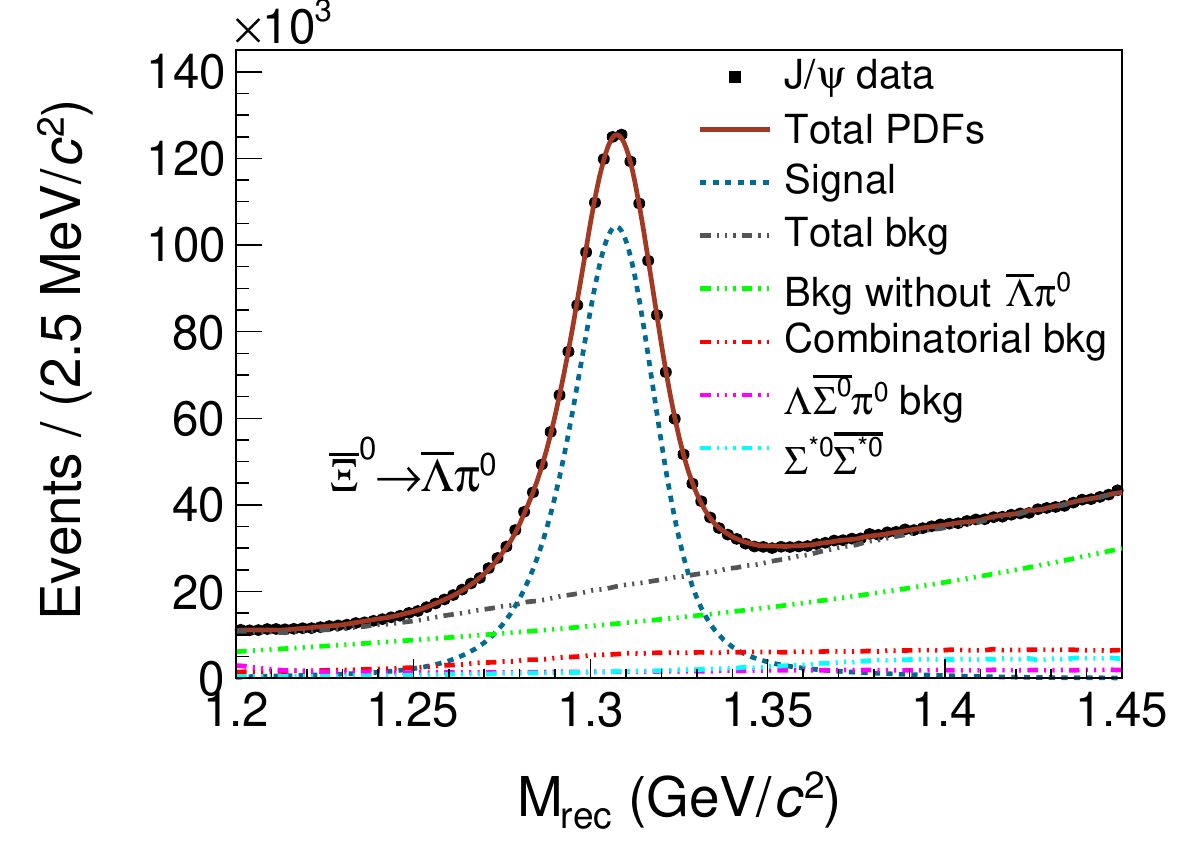}
        \put(25,55){(a)}
    \end{overpic}
    \begin{overpic}[width=0.48\textwidth]{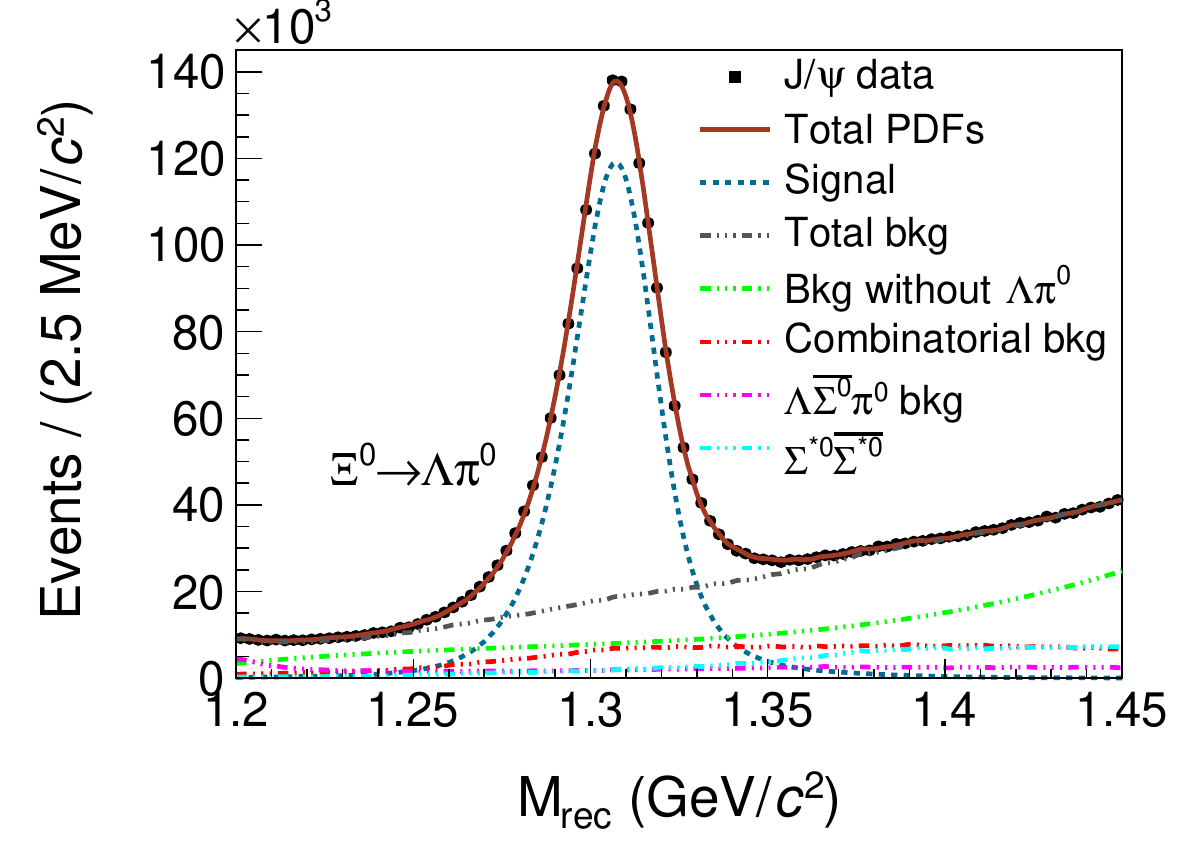}
        \put(25,55){(b)}
    \end{overpic}
    \caption{The fits to the $\rm{M_{rec}}$ distributions of (a) the $\bar\Xi^0$ tag and (b) the$\Xi^0$ ST candidates. The black dots are data, the solid red lines are the fit results, the dotted blue lines are the signal shapes, the dashed green lines are the background processes without $\bar\Lambda\pi^0$, the dashed red lines are the combinatorial backgrounds, the dashed pink lines are the $\Lambda\bar\Sigma^0\pi^0$ backgrounds, the dashed cyan lines are the $\Sigma^{*0}\bar\Sigma^{*0}$ backgrounds and the dashed grey lines are the total background shapes.}
    \label{fig:styield}
\end{figure}

The backgrounds of $J / \psi \rightarrow \pi^0 \Lambda \, \bar{\Sigma}^0+$ c.c., $J / \psi \rightarrow \Sigma^{0 *} \, \bar{\Sigma}^{0 *}$, and the combinatorial background of the signal are described by the shapes of the corresponding MC-simulated samples. Other backgrounds are described with a third-order polynomial function. To compensate for the resolution difference between data and MC simulation, the signal shape is convolved with a Gaussian function. The fit curves are shown in figure~\ref{fig:styield}. The ST yields and the detection efficiencies evaluated with the corresponding signal MC samples are summarized in table~\ref{tab:STyield}.  The ST efficiency correction factor is obtained by combining the correction factors for the detection efficiencies of $\Lambda, \bar{\Lambda}, \pi^\pm, \pi^0$, which are studied with the control samples $J/\psi \to \Xi^-\bar{\Xi}^+ \to (\Lambda \pi^-)(\bar{\Lambda}\pi^+)$ and $J/\psi \to \Xi^0\bar{\Xi}^= \to (\Lambda \pi^0)(\bar{\Lambda}\pi^0)$. 
The branching fraction of $J/\psi\to\Xi^0 \, \bar\Xi^0$ is calculated according to
\begin{equation}
    \mathcal{B}(J/\psi\to \Xi^0 \, \bar\Xi^0)=\frac{N_{\rm{ST}}}{N_{J/\psi} \, \mathcal{B}(\bar\Xi^0\to\bar\Lambda\pi^0) \, \mathcal{B}(\bar\Lambda\to\bar p\pi^+) \, \varepsilon_{\rm{ST}}},
\end{equation}
where $N_{\rm{ST}}$ is the ST yield, $N_{J/\psi}$ is the total number of $J/\psi$ events, and $\varepsilon_{\rm{ST}}$ is the ST detection efficiency. The systematic uncertainties of the branching fraction measurement of $\mathcal{B}(J/\psi\to\Xi^0 \, \bar\Xi^0)$ are dominated by the fit process ($N_{\rm{ST}}$), as outlined in section~\ref{sec:sys}. Other systematic uncertainties are from $N_{J/\psi},\ \mathcal{B}(\bar\Xi^0\to\bar\Lambda\pi^0),\ \mathcal{B}(\bar\Lambda\to\bar p\pi^+)$, and selection efficiency difference between data and MC simulation. The measured $\mathcal{B}(J/\psi\to\Xi^0 \, \bar\Xi^0)$  deviates about 2$\sigma$ from the previous measurement~\cite{jsixixib}.

\begin{table}[htbp]
    \centering
    \begin{tabular}{c|c|c}
        \hline
             & $\bar\Xi^0\to\bar\Lambda\pi^0$ mode & $\Xi^0\to\Lambda\pi^0$ mode \\
        \hline
        $N_{\rm{ST}} \, (\times10^3)$ & $1401.8\pm2.1$ & $1611.5\pm2.2$  \\
        $\varepsilon_{\rm{ST}}(\%)$ & $17.61$ & $19.78$ \\
        $\mathcal{B} \, (\times 10^{-3})$ & $1.241\pm 0.002$ & $1.270\pm 0.002$ \\
        Correction factor~\cite{BAM760} & $0.982$ & $1.006$ \\
        $\mathcal{B}_{\text{corr}} \, (\times 10^{-3})$ & $1.264\pm0.002\pm0.024$ & $1.262\pm0.002\pm0.030$ \\
        \hline
    \end{tabular}
    \caption{The ST yields, ST efficiencies and the obtained branching fractions, where the uncertainties are statistical only. The correction factor denotes the difference between the MC and data efficiencies in the ST event selection. $\mathcal{B}_{\text{corr}}$ is the corrected branch fraction, where the first and second uncertainties are statistical and systematic, respectively.}
    \label{tab:STyield}
\end{table}

On the signal side, to further suppress potential background events and improve the mass resolution, a seven-constraint (7C) kinematic fit is performed, constraining the total reconstructed four momentum to that of the initial $e^+e^-$ state and constraining $\gamma\gamma,\ \Lambda\gamma\gamma,\ \bar\Lambda\pi^0$ masses to the nominal masses of the $\pi^{0}$, $\Xi^{0}$ and $\bar\Xi^0$, respectively.  The event with minimum $\chi^2_{\rm{7C}}$ of the kinematic fit is selected, considering all additional photon pairs and all $\Lambda$ candidates. Events with $\chi^2_{\rm{7C}} < 50$ are kept. The requirement is optimized by a figure-of-merit, defined as $\frac S {\sqrt{S+B}}$ , where $S$ is the number of signal events and $B$ is the number of background events, estimated based on the MC simulation. The $\gamma\Lambda$ combination with minimum mass difference with the $\Sigma^0$ nominal mass is chosen as a $\Sigma^0$ candidate, leaving the other photon for the reconstruction of the $\Xi^0 \to \gamma \Sigma^0$ transition. 

To investigate other possible background processes, an inclusive MC sample of 10 billion $J/\psi$ events is examined by TopoAna~\cite{topana}, a software tool to categorize backgrounds and identify the physics processes of interest from the inclusive MC sample. The dominant background contributions can be classified into two main categories: I. $J/\psi\to \Xi^0 \, \bar\Xi^0\to \Lambda\pi^0 \, \bar\Lambda\pi^0$; II. $J/\psi\to\Sigma^{(*)0} \,\bar\Sigma^{(*)0}$ associated background, such as $J/\psi\to\Sigma^{*0} \, \bar\Sigma^{*0}, \ J/\psi\to \Sigma^0 \, \bar\Sigma^{*0},\ J/\psi\to \pi^0 \Sigma^0 \, \bar\Sigma^0$. For background I, the invariant mass of the two final state $\gamma$s is used to suppress it. The mass distribution of background I is expected to have a peak at $M_{\pi^0}$, while the signal process has a smooth distribution. Events with $88 <|M_{\gamma\gamma}|< 162\ \mathrm{MeV}/{\it c}^2$ are excluded based on optimizing a figure-of-merit,. For background II, the products from the $\Sigma^{(*)0}$ decay mostly originate from the interaction point, while the products from the $\Xi^0$ are often displaced from the interaction point due to the large decay length of the $\Xi^0$, $8.71\ \mathrm {cm}$. Thus, a secondary vertex fit is performed on the $\Lambda$ and $\bar\Lambda$ pair to get the decay length of a virtual particle, which decays to $\Lambda\bar\Lambda$, and events passing the secondary vertex fit with $\left | L/\sigma_L\right|<2.0$ are excluded. 

%Figure~\ref{fig:step} shows the data, inclusive MC and signal MC sample's invariant mass distribution of $\Sigma^0$ candidates after further event selection, which is the variable used to extract the yields of signal.

%\begin{figure}[htbp]
%    \centering
%    \includegraphics[width=0.48\textwidth]{plot/06DTana/3_stack.pdf}      
%    \caption{The $\gamma\Lambda$ invariant mass distributions of $\Sigma^0$ candidates after further selection. The black dots are data sample, blue line is background from inclusive MC and the red line is signal distribution. }
%    \label{fig:step}
%\end{figure}
\section{Branching fraction and decay asymmetry measurement}
\label{sec:result}
\begin{figure}[htbp]
    \centering
    \begin{overpic}[width=0.45\textwidth]{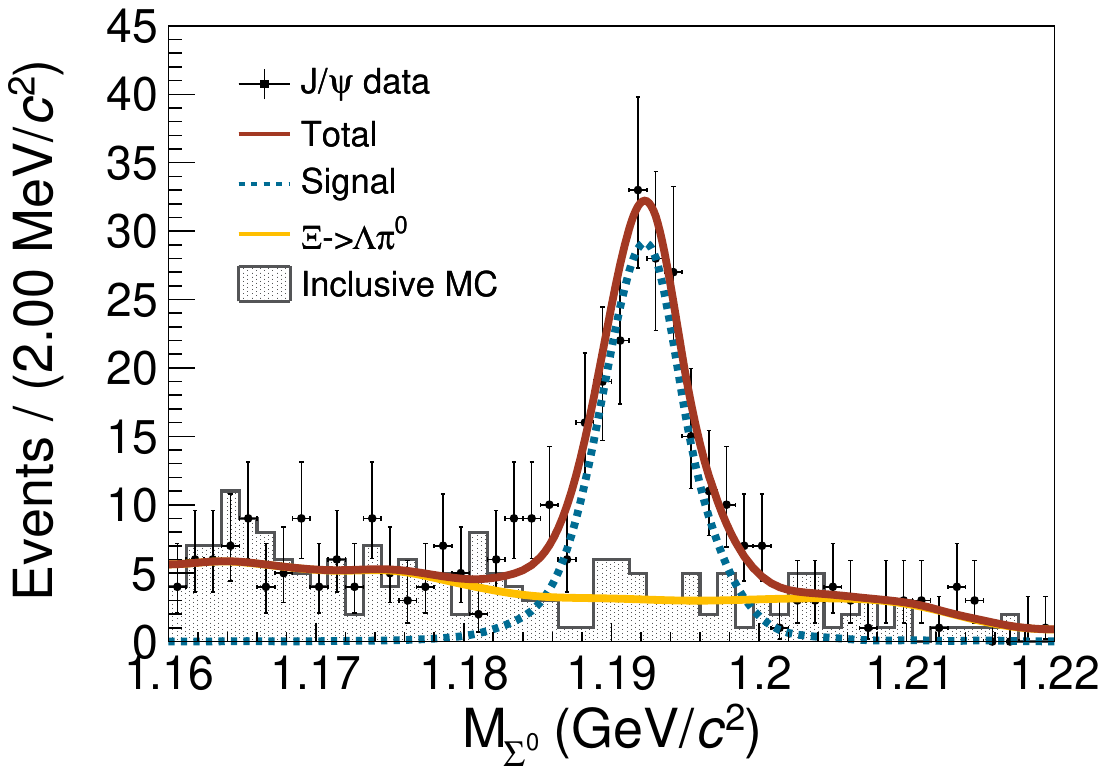}
        \put(75,55){(a)}
    \end{overpic}
    \begin{overpic}[width=0.45\textwidth]{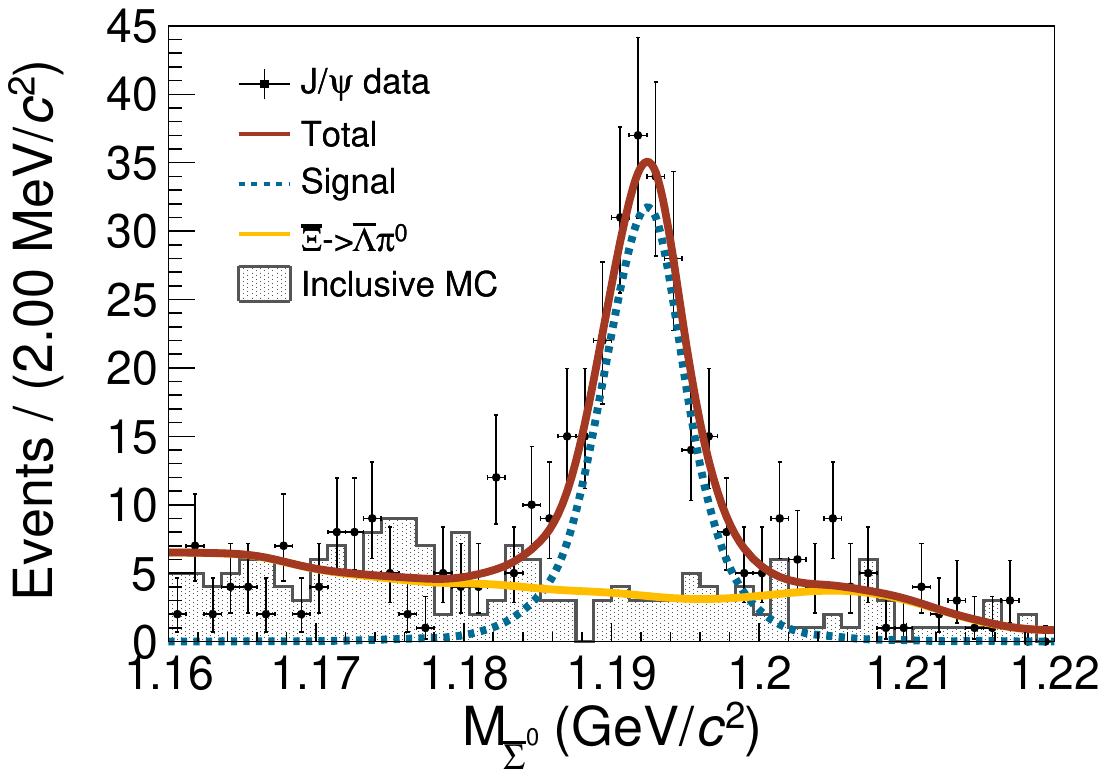}
        \put(75,55){(b)}
    \end{overpic}
    \caption{The simultaneous fit on the $M_{\Sigma^0(\bar\Sigma^0)}$ distributions for process of (a) $\Xi^0\to\gamma\Sigma^0$ and (b) $\bar\Xi^0\to\gamma\bar\Sigma^0$. The red solid lines are the total fit results, the blue dotted lines are signal components, the orange solid lines are background components, and the grey histograms represent the inclusive MC.}
    \label{DTfit}
\end{figure}

With the above selection criteria, there are significant enhancements in the $\gamma\Lambda$ mass spectrum close to the $\Sigma^0$ mass. To obtain the number of signal events, an unbinned maximum likelihood fit is performed on the $M_{\gamma\Lambda}$ distribution. The signal and background shapes are described by the corresponding exclusive MC samples, respectively. The branching fraction of $\Xi^0\to\gamma\Sigma^0$ is calculated according to
\begin{equation}
    \mathcal{B}(\Xi^0\to \gamma\Sigma^0) = \frac{N_{\rm{DT}}}{N_{\rm{ST}}} \, \frac{\varepsilon_{\rm{ST}}}{\varepsilon_{\rm{DT}}} \, \frac 1 {\mathcal{B}(\Lambda\to p\pi^-) \mathcal{B}(\Sigma^0\to \gamma\Lambda)},
    \label{eq:BF}
\end{equation}
where $N_{\rm{DT}}$ is the DT yield, $N_{\rm{ST}}$ is the ST yield, and $\varepsilon_{\rm{ST}}, \varepsilon_{\rm{DT}}$ are the ST and DT detection efficiencies, respectively, for each ST mode. The simultaneous fit results, which constrain the branching fractions of the two c.c.~channels to be equal, are shown in figure~\ref{DTfit}. The independent and simultaneous fit results are summarized in table~\ref{tab:DTyield}.

\begin{table}[htbp]
    \centering
    \begin{tabular}{c|c|c}
        \hline
        Decay & $\Xi^0\to\gamma\Sigma^0$ & $\bar\Xi^0\to\gamma\bar\Sigma^0$ \\
        \hline
        $N_{\rm {DT}}$ & $178\pm16$ & $214\pm17$  \\
        $\varepsilon_{\rm{DT}}~(\%)$ & $1.01$ & $1.04$ \\
        Individual $\mathcal{B}~(\times 10^{-3})$ & $3.45\pm 0.29$ & $3.95\pm 0.29$ \\
        Correction factor & $1.018$ & $0.991$\\ 
        Corrected individual $\mathcal{B}~(\times 10^{-3})$ & $3.39\pm0.28$ & $3.98\pm 0.29$\\
        \cline{2-3}
        Simultaneous $\mathcal{B}~(\times 10^{-3})$ & \multicolumn{2}{c}{$3.71\pm0.22$}\\
        Corrected simultaneous $\mathcal{B}~(\times 10^{-3})$ &\multicolumn{2}{c}{$3.69\pm0.21$}\\
        \hline
    \end{tabular}
    \caption{The summaries of DT yields, where the uncertainties are statistical only. The correction factor denotes the difference between the MC and data efficiencies in the DT event selection.}
    \label{tab:DTyield}
\end{table}

The decay asymmetry parameter is measured for the $\Xi^0 \to\gamma \Sigma^0$ and $\bar{\Xi}^0 \to \gamma\bar{\Sigma}^0 $ decay channels in the signal region: $1.178 < M_{\Sigma^0(\bar\Sigma^0)}<1.202 \ \mathrm{GeV}/{\it c}^2$. A joint likelihood function, $\mathcal{L}$, is constructed according to eq.~\ref{eq:decay_amplitude}; it incorporates the set of observables  $\xi=(\theta_{\Xi^0},\ \phi_{\Sigma^0},\ \theta_{\Sigma^0},\ \phi_{\Lambda},\ \theta_{\Lambda},\ \phi_{p},\ \theta_{p},\ \theta_{\bar\Lambda},\ \phi_{\bar\Lambda},\ \theta_{\bar p},\ \phi_{\bar p})$ and a set of characterized decay parameters $H=(\alpha_{J/\psi},\ \Delta\Phi,\ \alpha_{\Xi^0\to\gamma\Sigma^0}, \alpha_{\Lambda\to p\pi},\ \alpha_{\bar\Xi^0\to\bar\Lambda\pi^0},\ \Delta\Phi_{\bar\Xi\to\bar\Lambda\pi^0},\ \alpha_{\bar\Lambda\to\bar p \pi})$. The effect of detection efficiency on $\mathcal{L}$ is evaluated by a normalization factor, $\mathcal{N}$, calculated with the efficiency-corrected signal MC sample with the importance-sampling method. The target function for the fit is $-\ln \mathcal{L} =-\ln \mathcal{L}_{\text {data}}+\ln \mathcal{L}_{\text {bkg }}$, where the contributions from the background, $\mathcal{L}_{\text {bkg }}$, are estimated from the corresponding exclusive MC sample. The background likelihoods are normalized to the fitted yield numbers of the background events in the data sample. The fits are performed for the two decay channels individually. Furthermore, a simultaneous fit is also performed, assuming the same magnitude but opposite sign for the decay asymmetry parameters between the c.c. channels. The results are shown in table~\ref{tab:DecayAsy}. 

\begin{figure}[htbp]
    \centering
    \begin{overpic}[width=0.4\textwidth]{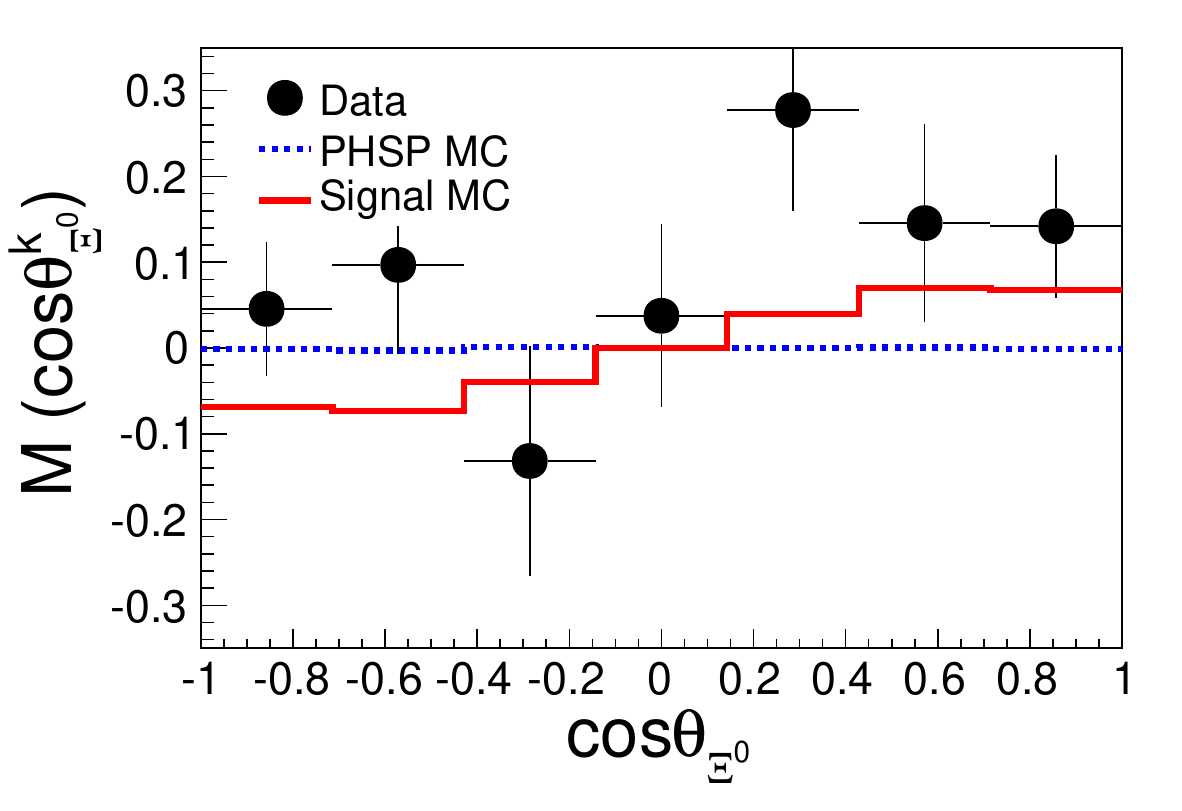}
        \put(25,25){(a)}
    \end{overpic}
    \begin{overpic}[width=0.4\textwidth]{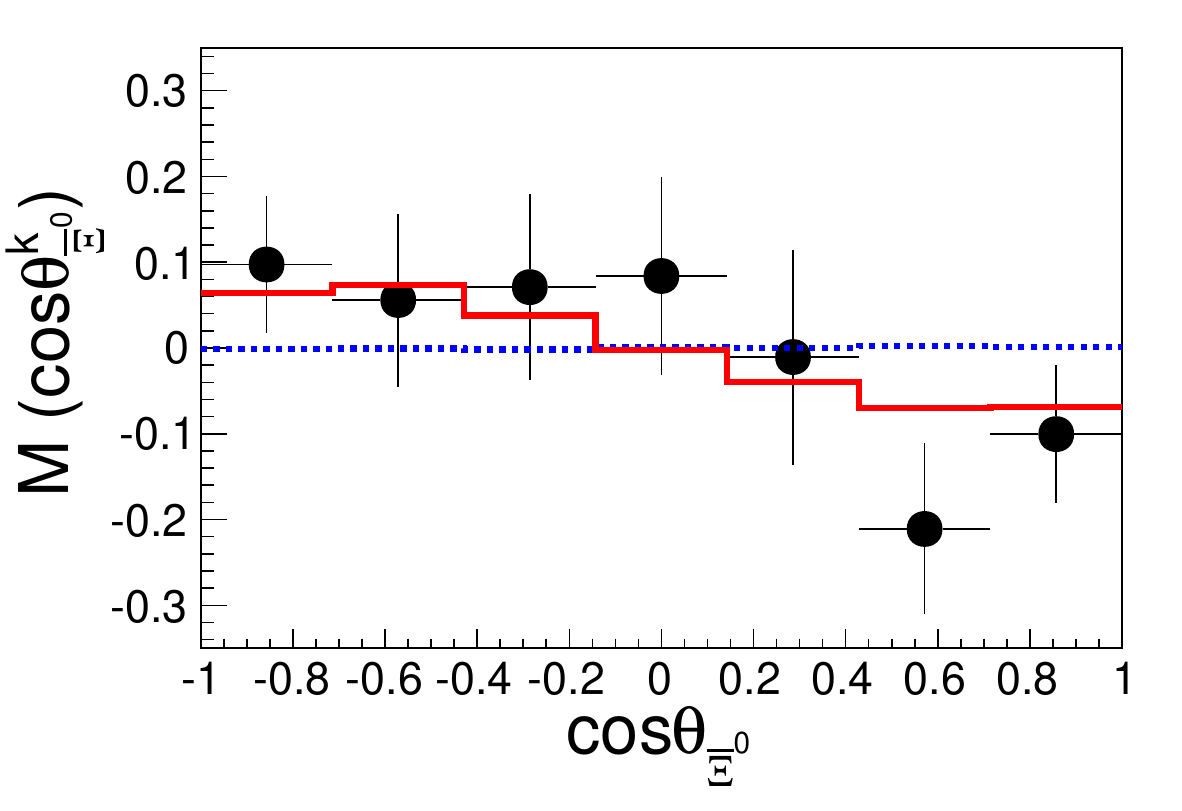}
        \put(25,25){(b)}
    \end{overpic}
    \caption{Moment distributions for (a) $\Xi^0\to\gamma\Sigma^0$ and (b) $\bar\Xi^0\to\gamma\bar\Sigma^0$.
      The black dots with error bars are data, the solid red lines are from signal MC, and the dashed blue lines
      are from phase-space MC.}
    \label{fig:moments}
\end{figure}

The effect of the decay asymmetry is visualized via moments, which is calculated for $m=7$ intervals in $\cos\theta_{\Xi^0}$:
\begin{equation}
    M(\cos\theta_{\Xi^0}^k)=\frac m N \sum_{i=1}^{N_k}\sin\theta_{\Sigma^0}^i\sin\phi_{\Sigma^0}^i,
\end{equation}
where $N$ is the total number of events, $\cos \theta_{\Xi^0}^k$ is the $k$-th interval of $\cos \theta_{\Xi^0}$, and $N_k$ is the number of events in the interval. Figure~\ref{fig:moments} shows the comparisons of moments between data and MC simulation.

%\begin{equation}
%    \mathcal{W}=
%    \sum_{\mu,\ \nu=0}^{3}\                    
%    \sum_{\mu',\ \nu',\ \rho=0}^3                     
%    C_{\mu \nu}             
%    b_{\mu \mu'}^{\Xi}      
%    a_{\nu \nu'}^{\bar{\Xi}} 
%    b_{\mu'\rho}^{\Sigma} 
%    a_{\nu'0}^{\bar\Lambda}  
%    a_{\rho0}^{\Lambda}
%    \label{eq:decay_amplitude}
%\end{equation}

\begin{table}[htbp]
    \centering
    \begin{tabular}{c|c|c}
        \hline
        Decay asymmetry & $\Xi^0\to \gamma\Sigma^0$ & $\bar\Xi^0\to\gamma\bar\Sigma^0$ \\
        \hline
        Individual fits & $-0.693\pm0.139_{\mathrm{stat}}$ & $0.956\pm0.139_\mathrm{stat}$ \\
        \cline{2-3}
        Simultaneous fit & \multicolumn{2}{c}{$-0.807\pm0.095_{\mathrm{stat}}$} \\
        \hline
    \end{tabular}
    \caption{The fit results for the decay asymmetry $\alpha_{\gamma}$.}
    \label{tab:DecayAsy}
\end{table}

%\begin{table}[htbp]
%    \centering
%    \begin{tabular}{c|c|c}
%    \hline
%    \hline
%    Decay mode  & $\Xi^0\to \Sigma^0\gamma$ & $\bar\Xi^0\to\bar\Sigma^0\gamma$ \\
%    \hline
%    $N_{DT}$    & $178\pm16$                & $214\pm17$                        \\
%    $\varepsilon_{DT}(\%)$ & $1.01$         & $1.04$                            \\
%    $\mathcal{B}(\times 10^{-3})$ & $3.39\pm0.28$ & $3.98\pm 0.29$              \\
%                                  & \multicolumn{2}{c}{$\mathbf{3.69\pm0.21}$}           \\
%    $\alpha_{\gamma}$ & $-0.693\pm0.139_{\mathrm{stat}}$ & $0.956\pm0.139_\mathrm{stat}$ \\
%                    &   \multicolumn{2}{c}{$\mathbf{-0.807\pm0.095_{\mathrm{stat}}}$} \\
%    \hline
%    \hline
%    \end{tabular}
%    \caption{The results of fits for the decays $\Xi^0\to \Sigma^0\gamma$ and $\bar\Xi^0\to\bar\Sigma^0\gamma$. The $\mathcal{B}$ and $\alpha_{\gamma}$ values are given both for individual and simultaneous fits. The first (second) uncertainties are statistical (systematic).}
%    \label{tab:sumresults}
%\end{table}
\section{Systematic uncertainty}
\label{sec:sys}
The systematic uncertainties on the $\mathcal B$ and decay asymmetry parameter measurements are considered separately. The uncertainties caused by ST event selection efficiencies are negligible because they cancel in the DT method. The uncertainties from ST yields are discussed in the analysis of Ref.~\cite{BAM760}, which has the same fit method as this analysis. The uncertainties of the $\mathcal B$ and decay asymmetry parameters are summarized in table~\ref{tab:sysBF} and table~\ref{tab:sysAsym}, respectively. In table~\ref{tab:sysBF}, the combined uncertainties associated with selection and ST yield are obtained by averaging the two c.c.~modes. In table~\ref{tab:sysAsym}, the fitting-related uncertainties are obtained from an input and output (IO) check via pull distributions. The total uncertainties are the sum of the quadrature of each term, assuming that these sources are independent. In the following, the $\Xi^0\to \gamma\Sigma^0$ decay mode is taken as an example to explain the calculation of systematic uncertainties.

\subsection{Systematic Uncertainties of $\mathcal{B}$ Measurement}
%\paragraph{ST yields related uncertainties}

\paragraph{Detection efficiency uncertainties.}
The uncertainties of $\pi^{\pm},\ \Lambda,$ and $\ \bar\Lambda$ detection efficiencies are taken from ref.~\cite{BAM760}. The corresponding uncertainties for protons and photons are from ref.~\cite{Sys_proton} and ref.~\cite{Sys_photon}, respectively. The individual uncertainties are calculated by weighting the corresponding uncertainties according to the signal distributions.

\paragraph{ST yields related uncertainties.}
In this analysis, the ST yields are extracted by performing binned maximum-likelihood fits to the $M_{\text{rec}}$ distributions, and the corresponding uncertainties include those associated with the fit range, the bin width, and the signal shapes, self-background shape, and the combinatorial background shape used in the fit.
\begin{itemize}
    \item  The systematic uncertainty associated with the fit range is obtained by performing the fits with alternative fit ranges, i.e., moving up or down in mass by $20\ \mathrm{MeV}/{\it c}^2$. The largest change, $0.5\%$, is considered as the uncertainty.
    \item To estimate the systematic uncertainty associated with the bin width in the binned fit, alternative fits are performed with the different widths. The largest difference, $0.5\%$, between the obtained yields is taken as the uncertainty.
    \item The uncertainties associated with the signal shape are estimated by performing alternative fits with the different means and widths of the convolved Gaussian function, which are varied with the residual values generated according to their uncertainties and correlation from the fit. The fit is repeated 500 times. The uncertainty of 0.2\% is estimated from the standard deviation of ST yields.
    %\begin{figure}
    %    \centering
    %    \includegraphics[width=0.5\textwidth]{plot/08SysUncertainty/sys_ST.png}
    %    \caption{Fitted ST yield distribution under different Gaussian parameters.}
    %    \label{fig:sys_ST}
    %\end{figure}
    \item The uncertainties associated with the self-background shape are estimated with an approach similar to the previous case, i.e., varying the parameters of the Gaussian function within their uncertainties, and the resultant standard deviation of 0.2\% on ST yields is taken as the uncertainty.
    \item In the nominal fit, the continuum background is described by a third-order Chebyshev polynomial function. Alternative fits with second- or fourth-order polynomial functions are performed, and the largest change in the signal yield, $1.5\%$, is taken as the uncertainty.
    \item The systematic uncertainty associated with the background shapes, which contains $\bar\Lambda\pi^0$, i.e., $J/\psi\to\Lambda\bar\Sigma^0\pi^0$ and $J/\psi\to\Sigma^{*0}\bar\Sigma^{*0}$ decays, is estimated by removing the corresponding components from the nominal fit model, individually. The largest change of the signal yield is found to be $0.4\%$, which is assigned as the uncertainty.
\end{itemize}

\paragraph{Event-level selection uncertainties.}
The uncertainties associated with the event-level requirements are studied with the control sample of $J/\psi\to\Xi^0\bar\Xi^0\to\Lambda\pi^0\bar\Lambda\pi^0$. The same selection criteria as for the signal process are used. The distributions of the kinematic fit $\chi^2$ and the decay length of particles, which decay to $\Lambda\bar\Lambda$, are examined. The resultant efficiency differences between the data and MC simulation are taken as systematic uncertainties. For the requirement for invariant mass of two photons, a new variable $M_{\gamma_{S}\gamma_{D}}$ is defined as the invariant mass of the two lower-energy photons from the $\pi^0$ on the ST and DT sides, respectively. A requirement of $ M_{\gamma_{S}\gamma_{D}} < 0.2\ \mathrm{GeV}/{\it c}^2$ is applied so that the distribution is almost the same as for $M_{\gamma\gamma}$ in the signal process. The efficiency of the two photon invariant mass requirement is calculated by $N_{M_{\gamma_{S}\gamma_{D}} \rm{requirement}} / N_{M_{\gamma_{S}\gamma_{D}}<0.2\ \mathrm{GeV}/{\it c}^2}$ and the difference between the data and MC simulation is taken as the systematic uncertainty.

\begin{figure}
    \begin{minipage}[t]{0.45\textwidth}
        \centering    
        \includegraphics[width=0.9\linewidth]{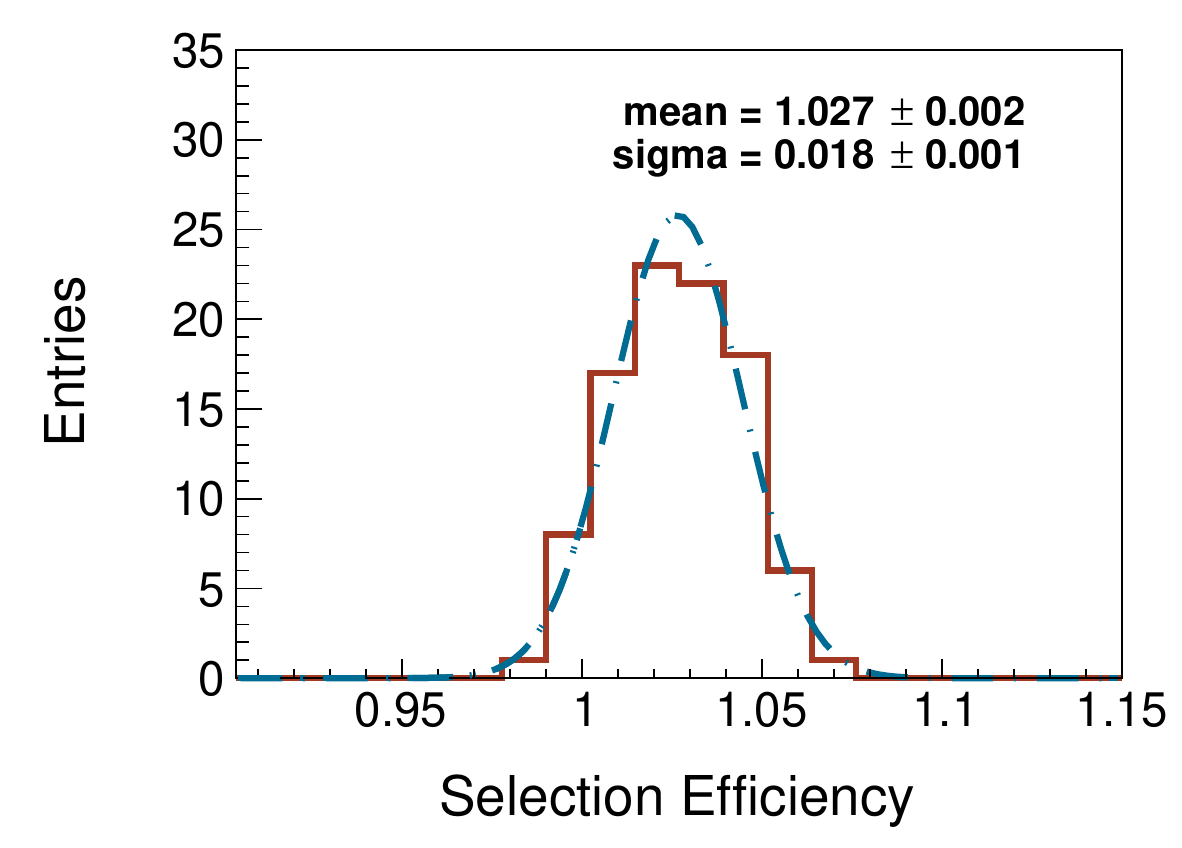}
    \end{minipage}
    \begin{minipage}[t]{0.45\textwidth}
        \centering    
        \includegraphics[width=0.9\linewidth]{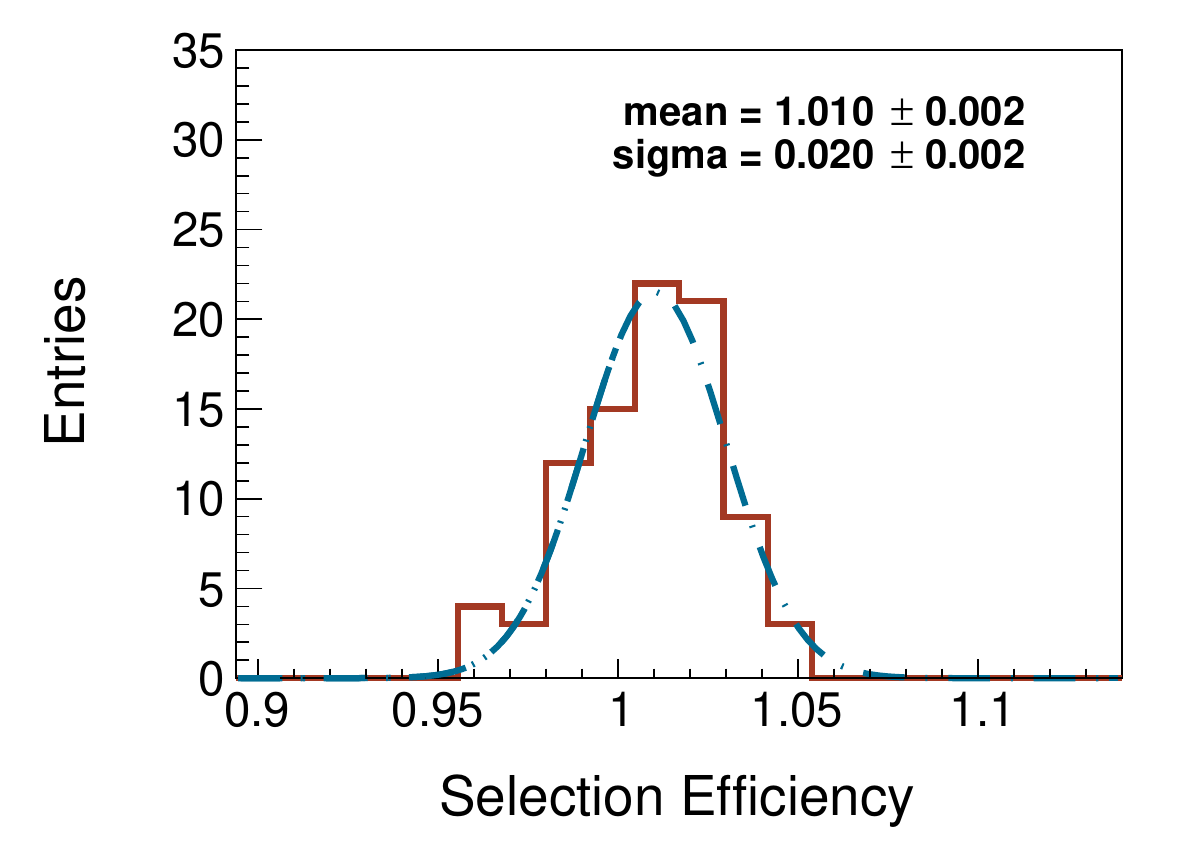}
    \end{minipage}
    \caption{The fitted distributions of selection efficiencies of 100 MC samples divided by the nominal values. The red histogram is the efficiency distribution and the blue line is the fitted result. The left plot is for the $\Xi^0\to\gamma\Sigma^0$ case and the right plot is for the c.c.~channel.}
    \label{fig:siguncer}
\end{figure}

\paragraph{Decay asymmetry parameters uncertainties.}
The uncertainties caused by the signal model are studied with 100 different MC samples that are produced with randomly generated decay asymmetry parameters according to their correlation matrix. The selection efficiencies are fitted with a Gaussian function, which is shown in figure~\ref{fig:siguncer}. The width of the Gaussian is taken as the systematic uncertainty.

\paragraph{DT yield fit uncertainties.}
Three sources of uncertainties from fit range, signal shape, and background shape in the DT yield fit are considered. The fit range is varied by $\pm 1\ \sigma$ of the signal width, $4\ \mathrm{MeV}/{\it c}^2$. The largest difference in DT yield is taken as the uncertainty. The signal shape is modified by convolving a Gaussian function with the MC shape, and the background shape is changed from exclusive MC shape to inclusive MC sample. The differences in the DT yields resulting from these changes are taken as the corresponding uncertainties.

\subsection{Systematic Uncertainty of Decay Asymmetry Measurement}

\paragraph{Detection efficiency uncertainties.}
The difference in efficiencies between the data and MC simulation for the proton, pion tracking and PID, photon reconstruction, and $\Lambda(\bar\Lambda)$ reconstruction will affect the normalization factor, which was used to normalize the likelihood function, and cause uncertainties for the measured decay asymmetry parameter. To estimate the corresponding uncertainties, the correction factors of these efficiencies are applied to the integrated MC sample to cancel out the difference in efficiency between data and MC simulation. The unbinned maximum likelihood fit is repeated, and the resultant differences of measured decay asymmetry parameters relative to the nominal values are taken as the uncertainties.

\paragraph{Event-level selection uncertainties.}
The uncertainties associated with the event-level requirements are studied with the control sample of $J/\psi\to\Xi^0\bar\Xi^0$. The same selection criteria as the signal process are used. The decay asymmetry of $\Xi^0\to\Lambda\pi^0$, which is used to mimic the decay of $\Xi^0\to \gamma\Sigma^0$ is measured with the same method as the signal process. Based on the selected sample, the event-level requirements are applied individually, and unbinned maximum likelihood fits are carried out. The results of $\alpha_{\Xi^0\to\Lambda\pi^0}$ are summarized in table~\ref{DecayEL}. The resultant differences with respect to the nominal value are taken as the uncertainties.

\begin{table}[htbp]
    \centering
    \begin{tabular}{c|c}
        \hline Source & $\Xi^0\to\Lambda\pi^0$\\
        \hline Nominal & $-0.3825$ \\
        \hline $\chi^2_{\rm{7C}}$ & $-0.3817$ \\
        \hline $M_{\gamma\gamma}$ & $-0.3782$ \\
        \hline $L/\sigma_{L}$ & $-0.3826$\\
        \hline
    \end{tabular}
    \caption{Summary of $\alpha_{\Xi^0\to\Lambda\pi^0}$ results with different event selection requirements.}
    \label{DecayEL}
\end{table}

\paragraph{Signal mass window.}
The systematic uncertainty associated with the fit range is obtained by performing several fits with alternate fit ranges~\cite{Barlow}, i.e. moving up or down by $2\ \mathrm{MeV}/{\it c}^2$ in mass several times on the control sample $J/\psi\to\Xi^0(\to\Lambda\pi^0) \, \bar\Xi^0(\to\bar\Lambda\pi^0)$.

%\begin{figure}[h]
%    \centering
%    \includegraphics[width=0.45\linewidth]{plot/08SysUncertainty/barlow.pdf}
%    \caption{Barlow test for signal mass window, the blue band is the statistical uncertainty, red dots are the deviation of fit result from nominal value, the red line is the linear fit result.}
%    \label{fig:barlow}
%\end{figure}

\begin{figure}[htbp]
    \centering
    \begin{overpic}[width=0.4\textwidth]{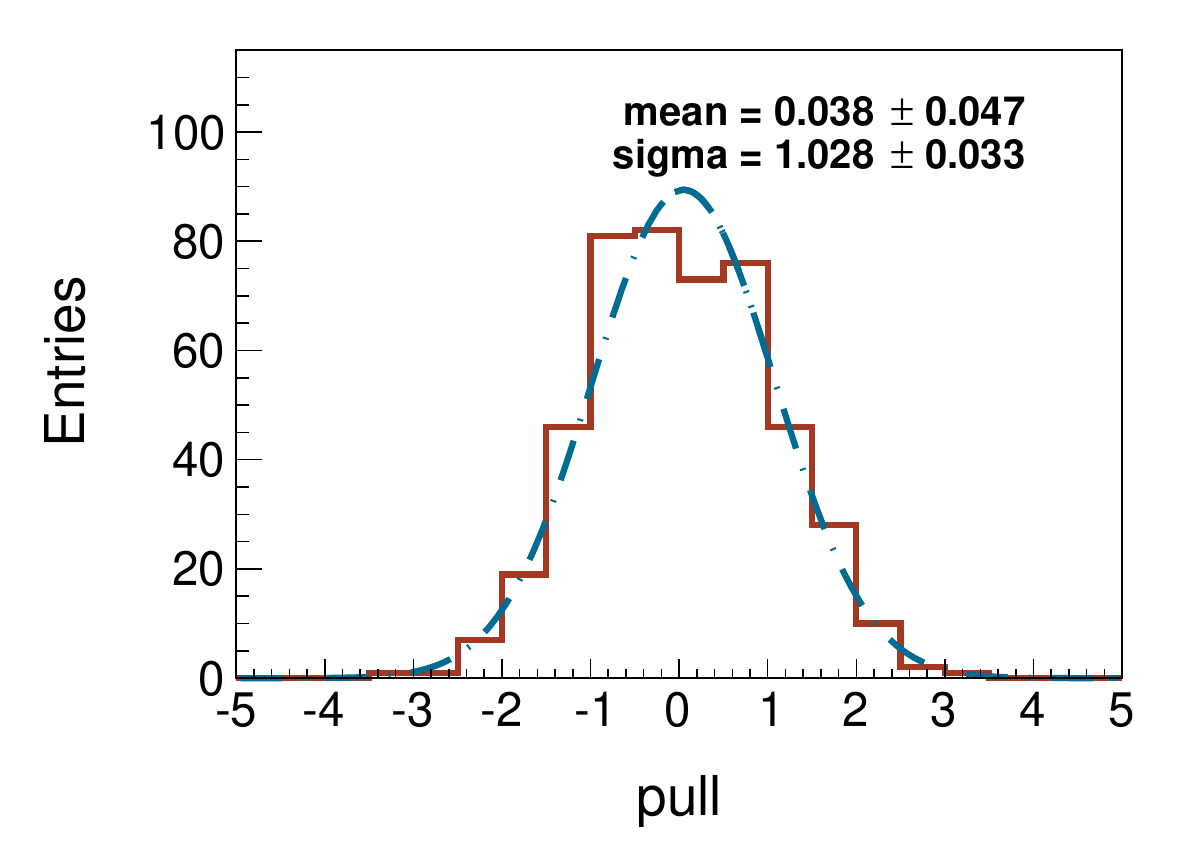}
        \put(75, 40){(a)}
    \end{overpic}
    \begin{overpic}[width=0.4\textwidth]{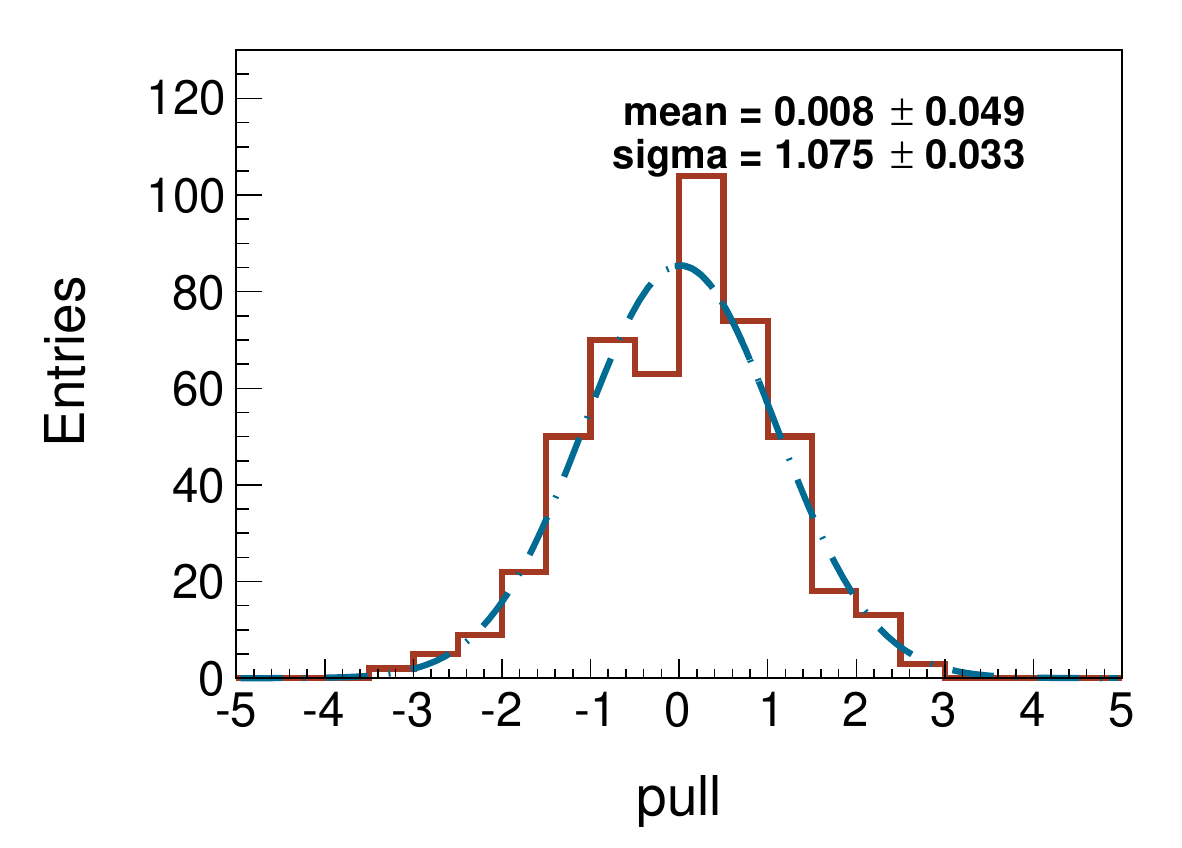}
        \put(75, 40){(b)}
    \end{overpic}
    \begin{overpic}[width=0.4\textwidth]{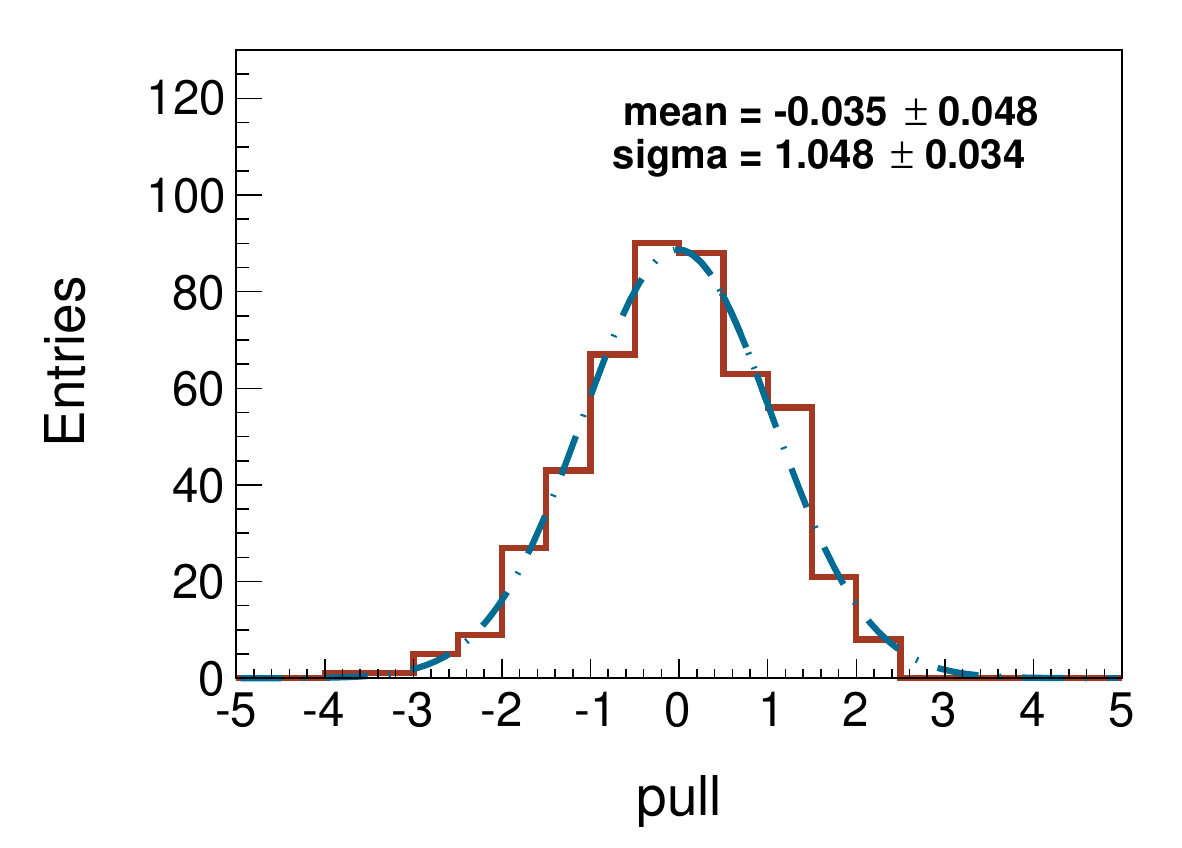}
        \put(75, 40){(c)}
    \end{overpic}
    \caption{The pull distribution of 500 independent fits for (a) $\Xi^0\to\gamma\Sigma^0$, (b) $\bar\Xi^0\to\gamma\bar\Sigma^0$ and (c) simultaneous fit.}
    \label{fig:Pull}
\end{figure}

\paragraph{Fit related uncertainties.}
The pull of the decay asymmetry parameter is $\text{pull} = (\alpha_{\rm{output}} - \alpha_{\rm{input}})/\sigma_{\alpha}$, where $\alpha_{\rm{output}}$ is the result from a fit, $\alpha_{\rm{input}}$ is the true decay asymmetry parameter, and $\sigma_{\alpha}$ is the statistical uncertainty of each fit. The systematic fitting uncertainties are estimated from the mean deviation plus the $1\sigma$ statistical error from 500 input and output checks, as shown in figure~\ref{fig:Pull}. 
 %According to definition of pull distribution Eq.~(\ref{eq:pull}), the uncertainties could be estimated by the mean deviation plus one times of its statistical error from the 500 times IO check, which is shown in figure~\ref{fig:Pull}, times corresponding statistical uncertainty.

\begin{table}[htbp]
    \centering
    \begin{tabular}{c|c|c|c}
        \hline Source & $\Xi^0\to \gamma\Sigma^0 \ (\%) $ & $\bar\Xi^0\to\gamma \bar\Sigma^0\ ( \%)$ & Combined (\%) \\
        \hline \multicolumn{4}{c}{ Selection efficiency related} \\
        \hline Proton track and PID & 0.1 & 0.3 & 0.2 \\
        \hline Pion track and PID & 0.2 & 0.2 & 0.2 \\
        \hline Lambda vertex fit & 0.2 & 0.1 & 0.2 \\
        \hline Photon detection & 0.2 & 0.2 & 0.2 \\
        \hline Decay length requirement& 0.3 & 0.2 & 0.3 \\
        \hline Kinematic fit $\chi^2$ requirement& 0.1 & 0.0 & 0.0 \\
        \hline $M_{\gamma\gamma}$ requirement & 0.6 & 0.3 & 0.5 \\
        \hline Decay parameters & 0.0 & 0.0 & 0.0 \\
        \hline \multicolumn{4}{c}{ ST yield fit }\\
        \hline Fit range & 0.5 & 0.4 & 0.5 \\
        \hline Bin width & 0.5 & 0.6 & 0.6 \\
        \hline Signal shape & 0.2 & 0.2 & 0.2 \\
        \hline Self-background shape & 0.2 & 0.7 & 0.5 \\
        \hline Combinatorial background shape & 1.5 & 1.9 & 1.7 \\
        \hline Bump-like background shape & 0.4 & 0.7 & 0.6 \\
        \hline \multicolumn{4}{c}{ DT yield fit } \\
        \hline Fit range & 1.7 & 1.9 & 1.0 \\
        \hline Signal shape & 1.1 & 1.4 & 1.8 \\
        \hline Background shape & 1.7 & 0.5 & 0.8 \\
        \hline $\mathcal{B}(\Lambda\to p\pi^-)$ & 0.8 & 0.8& 0.8\\
        \hline \multicolumn{4}{c}{ Summary } \\
        \hline Total & 3.3 & 3.4 & 3.2\\
        \hline
    \end{tabular}
    \caption{Systematic uncertainties of $\mathcal{B}$ measurements.}
    \label{tab:sysBF}
\end{table}

\begin{table}[htbp]
    \centering
    \begin{tabular}{c|c|c|c}
        \hline Source & $\Xi^0\to\gamma\Sigma^0$ & $\bar\Xi^0\to \gamma\bar\Sigma^0$ & Combined \\
        \hline \multicolumn{4}{c}{ Selection efficiency } \\
        \hline Track detection & 0.001 & 0.002 & 0.002 \\
        \hline Kinematic fit $\chi^2$ requirement & 0.001 & 0.004 & 0.002\\
        \hline $M_{\gamma\gamma}$ requirement& 0.004 & 0.007 & 0.005\\
        \hline Decay length & 0.000 & 0.004 & 0.002\\
        %\hline Signal mass window & 0.009 & 0.014 & 0.011\\
        \hline Signal mass window & 0.002 & 0.002 & 0.002\\
        \hline \multicolumn{4}{c}{ Fit related }\\
        \hline IO results & 0.012 & 0.008 & 0.008\\
        \hline \multicolumn{4}{c}{ Summary } \\
        \hline Total & 0.013 & 0.014 & 0.011\\
        \hline
    \end{tabular}
    \caption{Systematic uncertainties of decay asymmetry parameter measurements.}
    \label{tab:sysAsym}
\end{table}

\section{Summary}
\label{sec:sum}
\begin{figure}[htbp]
    \centering
    \includegraphics[width=0.55\textwidth]{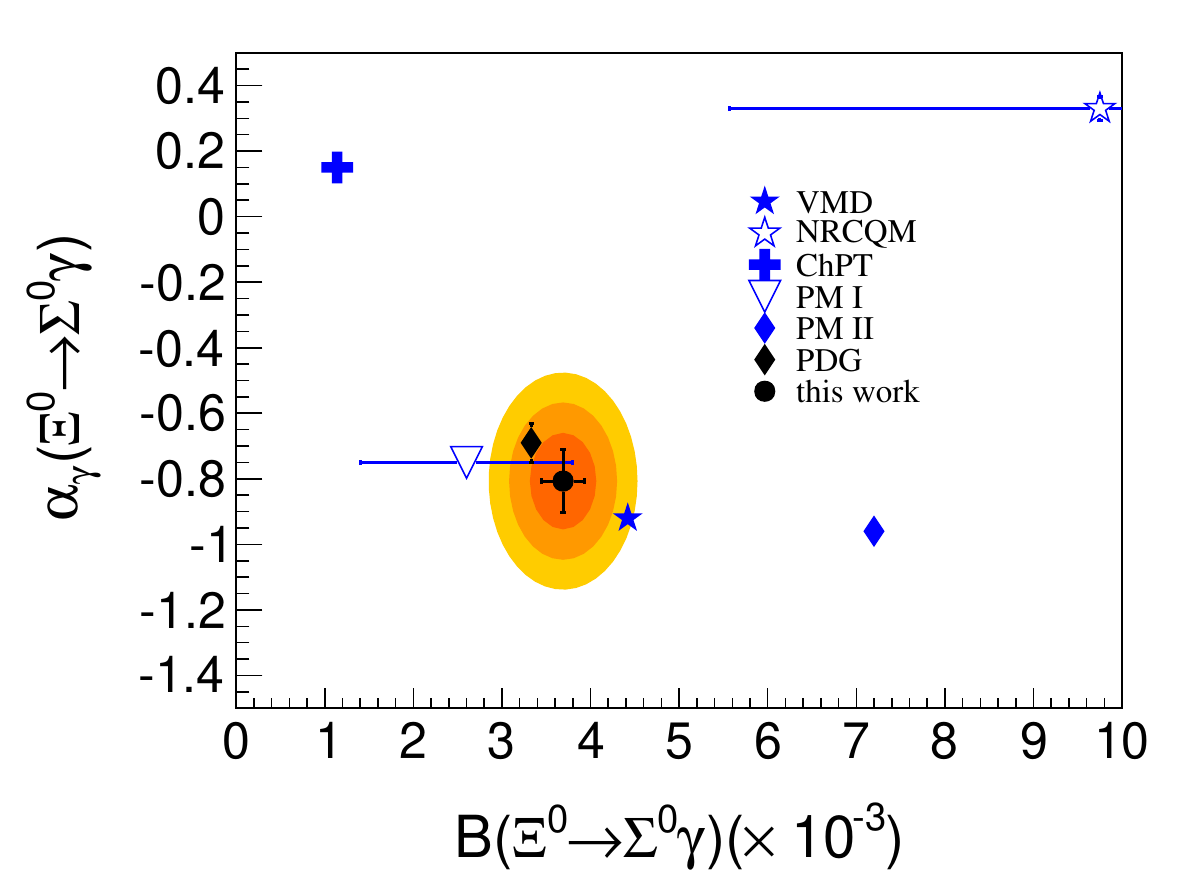}
    \caption{The distribution of $\mathcal{B}$ versus decay asymmetry of $\Xi^0\to\gamma\Sigma^0$. The black dot and diamond with error bars denote the BESIII result and the PDG value, respectively. The symbols in blue represent the results predicted in the VDM~\cite{PhysRevD.73.076005}, the NRCQM~\cite{CPCabc067}, the ChPT~\cite{PhysRevD.59.054019}, the PM I~\cite{NARDULLI1987187}, and PM II~\cite{GAVELA1981417}. The contours in orange gradient are the $68.2\%,\ 95.4\%$, and $99.7\%$ confidence level regions.}
    %\caption{Two dimensional distribution of the $\mathcal{B}$ and decay asymmetry of $\Xi^0\to\gamma\Sigma^0$. The black dot and diamond with error bars denote the BESIII result and the PDG value respectively. Other symbols in blue stand for the results predicted in the vector dominance model (VDM)~\cite{PhysRevD.73.076005}, the non-relativistic constituent quark model (NRCQM)~\cite{CPCabc067}, the chiral perturbation theory (ChPT)~\cite{PhysRevD.59.054019} and the pole models (PM I~\cite{NARDULLI1987187} and PM II~\cite{GAVELA1981417}). The contours in orange gradient represent $68.2\%,\ 95.4\%$, and $99.7\%$ confidence level of the $\mathcal{B}$ and $\alpha_{\gamma}$.}
    \label{fig:summary}
\end{figure}

The weak radiative decay $\Xi^0\to\gamma\Sigma^0$ is studied in the decay  $J/\psi\to\Xi^0 \, \bar\Xi^0$ by analyzing $(10.087\pm 0.044)\times 10^{9}$ $J/\psi$ events collected with the BESIII detector at BEPCII in 2009, 2012, 2018 and 2019.  The absolute branching fraction and the decay asymmetry parameters of $\Xi^0\to\gamma\Sigma^0$ are measured
and are summarized in table~\ref{Summary} and shown in figure~\ref{fig:summary}.

%Another thing worth noticing is that though the statistical uncertainty for decay asymmetry measurement is vary large, nearly $400$ events are used in the analysis. To reach the same level of statistical uncertainty with NA48 measurement in 2010~\cite{2010241}, which has about $15000$ events, only about $4000$ events are needed on the $e^+e^-$ collider experiment. The future $e^+e^-$ collider Super $\tau$-Charm facility(STCF)~\cite{STCF} is designed with a peak luminosity of $0.5\times 10^{35}\ \mathrm{cm}^{-2}\mathrm{s}^{-1}$ or higher. It will produce a data sample about a factor of 100 times larger than BEPCII, which is expected to realize much higher measurement accuracy.

%\begin{table}[htbp]
%    \centering
%    %\begin{tabular}{c|S[table-format=1.2(2)]|S[table-format=-1.3(3)]}
%    \begin{tabular}{c|c|c}
%        \hline
%        Decay & $\mathcal{B}$($\times 10 ^{-3}$) & $\alpha_{\gamma}$ \\
%        \hline
%        $\Xi^0\to\gamma\Sigma^0$ & $3.39\pm 0.28_{\text{stat}} \pm 0.11_{\text{syst}}$ & $-0.693\pm 0.139_{\text{stat}}\pm 0.013_{\text{syst}}$ \\
%        $\bar\Xi^0\to\gamma\bar\Sigma^0$ & $3.98\pm0.29_{\text{stat}} \pm 0.14_{\text{syst}}$ & $0.956\pm 0.139_{\text{stat}}\pm0.014_{\text{syst}}$\\
%        Combined & $3.69\pm0.21_{\text{stat}}\pm0.12_{\text{syst}}$ & $-0.807\pm0.095_{\text{stat}}\pm0.011_{\text{syst}}$\\
%        %\hline
%        %PDG value & $3.33\pm0.10$& $-0.69\pm 0.06$\\
%        \hline
%    \end{tabular}
%    \caption{$\mathcal{B}$ and $\alpha_{\gamma}$ results of this measurement.}
%    \label{Summary}
%\end{table}

\begin{table}[htbp]
    \centering
    \begin{tabular}{c|r@{\,$\pm$\,}l|r@{\,$\pm$\,}l}
        \hline
        Decay & \multicolumn{2}{c|}{$\mathcal{B}$ ($\times 10^{-3}$)} & \multicolumn{2}{c}{$\alpha_{\gamma}$} \\
        \hline
        $\Xi^0\to\gamma\Sigma^0$ & $3.39$ & $0.28_{\text{stat}}$ $\pm$ $0.11_{\text{syst}}$ & $-0.693$ & $0.139_{\text{stat}}$ $\pm$ $0.013_{\text{syst}}$ \\
        $\bar\Xi^0\to\gamma\bar\Sigma^0$ & $3.98$ & $0.29_{\text{stat}}$ $\pm$ $0.14_{\text{syst}}$ & $0.956$ & $0.139_{\text{stat}}$ $\pm$ $0.014_{\text{syst}}$ \\
        Combined & $3.69$ & $0.21_{\text{stat}}$ $\pm$ $0.12_{\text{syst}}$ & $-0.807$ & $0.095_{\text{stat}}$ $\pm$ $0.011_{\text{syst}}$ \\
        \hline
    \end{tabular}
    \caption{$\mathcal{B}$ and $\alpha_{\gamma}$ results of this measurement.}
    \label{Summary}
\end{table}

\begin{figure}[H]
    \centering
    \includegraphics[width=0.5\textwidth]{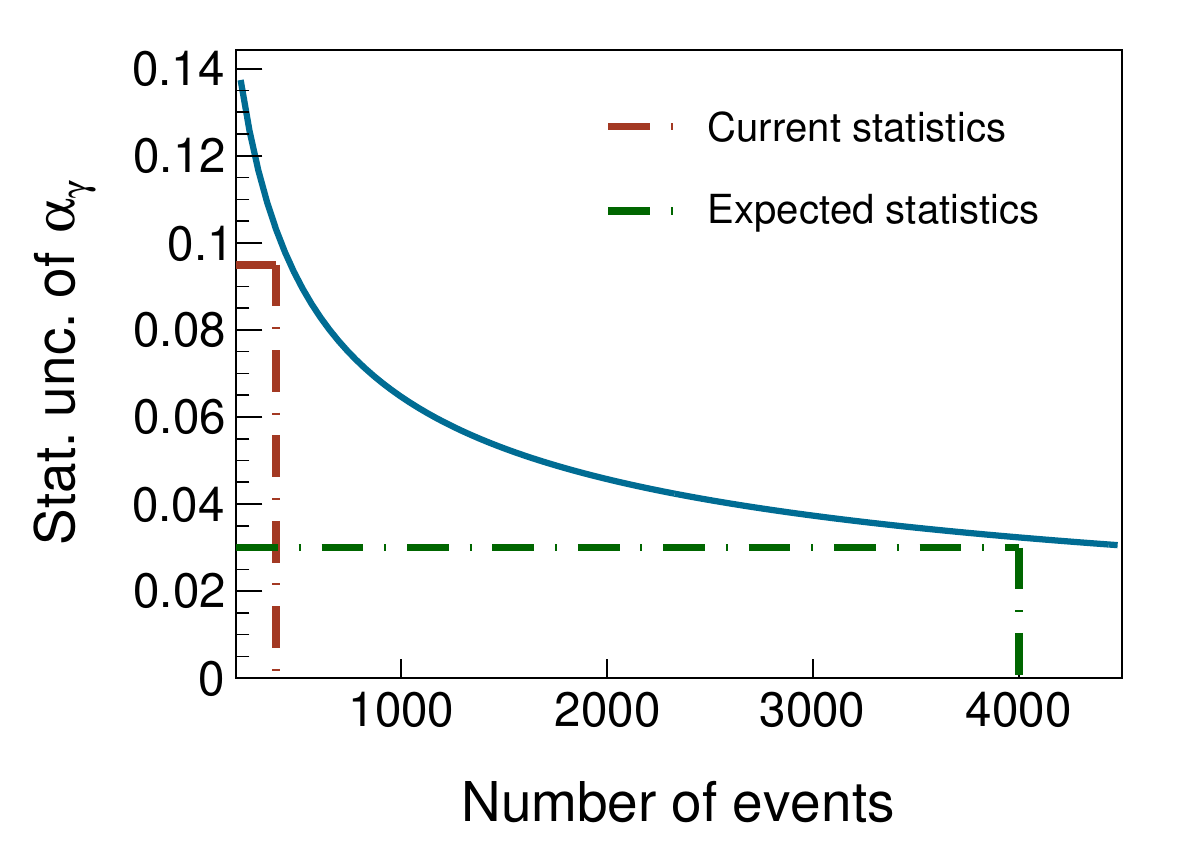}
    \caption{The statistical uncertainty of the decay asymmetry versus the number of signal events. The dashed red lines represent the current statistics and the corresponding statistical uncertainty. The dashed green lines represent the expected statistics for an $e^+e^-$ collider to reach the statistical uncertainty of the NA48 measurement.}
    \label{fig:error}
\end{figure}

The results are consistent with PDG value within $2\sigma$, favoring Pole model I~\cite{NARDULLI1987187} and the VMD model~\cite{PhysRevD.73.076005}. The current uncertainties are dominated by statistics. However, the study of HWRD shows great potential at $e^+e^-$ colliders.  The statistical uncertainty of the decay asymmetry parameter versus the number of signal events is shown in figure~\ref{fig:error}. To achieve a similar statistical uncertainty as the 2010 NA48 measurements~\cite{2010241}, about $4000$ events are needed at an $e^+e^-$ collider. This corresponds to only about a quarter of the number of events for a fixed target experiment. The future $e^+e^-$ collider Super $\tau$-Charm facility (STCF)~\cite{STCF} is designed with a peak luminosity of $0.5\times 10^{35}\ \mathrm{cm}^{-2}\mathrm{s}^{-1}$ or higher. It will produce a $J/\psi$ data sample about a factor of 100 times larger than BEPCII, leading to much higher measurement accuracies.

\acknowledgments

The BESIII Collaboration thanks the staff of BEPCII (https://cstr.cn/31109.02.BEPC) and the IHEP computing center for their strong support and the supercomputing center of USTC for their strong support. This work is supported in part by National Key R\&D Program of China under Contracts Nos. 2023YFA1609400, 2023YFA1606000, 2023YFA1606704; National Natural Science Foundation of China (NSFC) under Contracts Nos. 11635010, 11935015, 11935016, 11935018, 12025502, 12035009, 12035013, 12061131003, 12105276, 12122509, 12192260, 12192261, 12192262, 12192263, 12192264, 12192265, 12221005, 12225509, 12235017, 12361141819; the Chinese Academy of Sciences (CAS) Large-Scale Scientific Facility Program; CAS under Contract No. YSBR-101;Joint Large-Scale Scientific Facility Funds of the NSFC and CAS under Contracts Nos. U2032111; 100 Talents Program of CAS; CAS Youth Team Program under Contract No. YSBR-101; The Institute of Nuclear and Particle Physics (INPAC) and Shanghai Key Laboratory for Particle Physics and Cosmology; Agencia Nacional de Investigación y Desarrollo de Chile (ANID), Chile under Contract No. ANID PIA/APOYO AFB230003; German Research Foundation DFG under Contract No. FOR5327; Istituto Nazionale di Fisica Nucleare, Italy; Knut and Alice Wallenberg Foundation under Contracts Nos. 2021.0174, 2021.0299; Ministry of Development of Turkey under Contract No. DPT2006K-120470; National Research Foundation of Korea under Contract No. NRF-2022R1A2C1092335; National Science and Technology fund of Mongolia; Polish National Science Centre under Contract No. 2024/53/B/ST2/00975; Swedish Research Council under Contract No. 2019.04595; U. S. Department of Energy under Contract No. DE-FG02-05ER41374

% Bibliography

%% [A] Recommended: using JHEP.bst file
\bibliographystyle{JHEP}
%\bibliography{biblio.bib}

\providecommand{\href}[2]{#2}\begingroup\raggedright\endgroup
%% or
%% [B] Manual formatting (see below)
%% (i) We suggest to always provide author, title and journal data or doi:
%% in short all the informations that clearly identify a document.
%% (ii) please avoid comments such as "For a review'', "For some examples",
%% "and references therein" or move them in the text. In general, please leave only references in the bibliography and move all
%% accessory text in footnotes.
%% (iii) Also, please have only one work for each \bibitem.

%%\begin{thebibliography}{99}
%%
%%\bibitem{a}
%%Author,
%%\emph{Title},
%%\emph{J. Abbrev.} {\bf vol} (year) pg.
%%
%%\bibitem{b}
%%Author,
%%\emph{Title},
%%arxiv:1234.5678.
%%
%%\bibitem{c}
%%Author,
%%\emph{Title},
%%Publisher (year).
%%
%%\end{thebibliography}
\newpage
%\paragraph{The BESIII Collaboration}
\section*{The BESIII Collaboration}
%% Saved at => 2025-03-19
M.~Ablikim$^{1}$\BESIIIorcid{0000-0002-3935-619X},
M.~N.~Achasov$^{4,b}$\BESIIIorcid{0000-0002-9400-8622},
P.~Adlarson$^{77}$\BESIIIorcid{0000-0001-6280-3851},
X.~C.~Ai$^{82}$\BESIIIorcid{0000-0003-3856-2415},
R.~Aliberti$^{36}$\BESIIIorcid{0000-0003-3500-4012},
A.~Amoroso$^{76A,76C}$\BESIIIorcid{0000-0002-3095-8610},
Q.~An$^{73,59,\dagger}$,
Y.~Bai$^{58}$\BESIIIorcid{0000-0001-6593-5665},
O.~Bakina$^{37}$\BESIIIorcid{0009-0005-0719-7461},
Y.~Ban$^{47,g}$\BESIIIorcid{0000-0002-1912-0374}
% ...
\\
\vspace{0.2cm}
(BESIII Collaboration)\\
\vspace{0.2cm} {\it
$^{1}$ Institute of High Energy Physics, Beijing 100049, People's Republic of China\\
% ...
$^{4}$ Budker Institute of Nuclear Physics SB RAS (BINP), Novosibirsk 630090, Russia\\
% ...
$^{36}$ Johannes Gutenberg University of Mainz, Johann-Joachim-Becher-Weg 45, D-55099 Mainz, Germany\\
$^{37}$ Joint Institute for Nuclear Research, 141980 Dubna, Moscow region, Russia\\
% ...
$^{47}$ Peking University, Beijing 100871, People's Republic of China\\
% ...
$^{58}$ Southeast University, Nanjing 211100, People's Republic of China\\
$^{59}$ State Key Laboratory of Particle Detection and Electronics, Beijing 100049, Hefei 230026, People's Republic of China\\
% ...
$^{73}$ University of Science and Technology of China, Hefei 230026, People's Republic of China\\
% ...
$^{76}$ University of Turin and INFN, (A)University of Turin, I-10125, Turin, Italy; (B)University of Eastern Piedmont, I-15121, Alessandria, Italy; (C)INFN, I-10125, Turin, Italy\\
$^{77}$ Uppsala University, Box 516, SE-75120 Uppsala, Sweden\\
% ...
$^{82}$ Zhengzhou University, Zhengzhou 450001, People's Republic of China\\

\vspace{0.2cm}
$^{\dagger}$ Deceased\\
% ...
$^{b}$ Also at the Novosibirsk State University, Novosibirsk, 630090, Russia\\
% ...
$^{g}$ Also at State Key Laboratory of Nuclear Physics and Technology, Peking University, Beijing 100871, People's Republic of China\\
% ...

}
%% ends here %%

\end{document}